\documentclass[12pt]{article}
\usepackage{epsfig}
\usepackage{amssymb}
\newcommand{\be}{\begin{equation}}
\newcommand{\ee}{\end{equation}}
\newcommand{\bea}{\begin{eqnarray}}
\newcommand{\eea}{\end{eqnarray}}

\def\bmu {B_{\mu}}
\def\wmu {W^{3}_{\mu}}
\def\wmupm {W^{\pm}_{\mu}}

\def\bh {B_H}
\def\wh {W^{3}_{H}}
\def\whpm {W^{\pm}_{H}}

\begin{document}

\thispagestyle{empty}
\begin{flushright}

\end{flushright}
\vskip 15pt

\begin{center}
{\Large {\bf Probing two Universal Extra
    Dimension model with leptons and photons at the LHC and ILC}}
\renewcommand{\thefootnote}{\alph{footnote}}

\hspace*{\fill}

\hspace*{\fill}

{ \tt{
Kirtiman Ghosh\footnote{E-mail address:
kirtiman.ghosh@saha.ac.in}
}}\\

\small {\em Department of Physics, University of Calcutta,\\
92, A. P. C. Road, Kolkata 700009, India.}
\\

\vskip 40pt

{\bf ABSTRACT}
\end{center}

\vskip 0.5cm

We discuss the collider signatures of electroweak $(1,0)$-mode
excitations in the framework of two universal extra dimension
(2UED) at the LHC and ILC. In general, pair production of electroweak
$(1,0)$-mode 
particles of 2UED gives rise to {\em multi lepton plus missing transverse
momentum} signal. Upto 1/R = 400 GeV {\em 2-lepton plus missing
transverse momentum} signal and upto 1/R = 600 GeV {\em 3-lepton
plus missing transverse momentum} signal stands over the $5\sigma$
standard deviation of the Standard Model 
background at the LHC with $100~fb^{-1}$ integrated luminosity. At ILC
we study {\em single photon plus missing energy} signal from the
production of $U(1)$ gauge boson in association with a $U(1)$ spinless
adjoint. With proper choice of beam polarization, signal strength is
greater than  $5\sigma$ standard deviation of the Standard Model
background almost upto the kinematic limit of the collider.

\vskip 30pt

\section{Introduction}
One of the goals for the future collider experiments
will be to find out whether a new dynamics beyond the Standard Model
(SM) really exists around the TeV scale of energy. A great effort have
been also put in to pin down the exact nature of this new dynamics. In
this endeavor, supersymmetry and models with one or
more {\em extra dimension} play very special role.

Extra dimensional theories can be classified into several
classes. Models of ADD \cite{add} or RS \cite{rs} have been proposed to
circumvent the long-standing hierarchy problem. In this framework,
gravity lives in $(4+D)$ dimensions and the SM particles
are confined to a 3-brane (a $(3+1)$ dimensional space) embedded in
the 
$(4+D)$ dimensional bulk, with $D$ spatial dimensions compactified on
a volume $V$. For large enough value of this extra dimensional volume
$V$, the fundamental $4+D$ dimensional Plank mass can be as low as $1$
TeV, although the effective $4$ dimensional Plank mass can be as large
as $10^{19}$ GeV. There are some interesting generalization of these
models in which the SM
particles are confined to a $(3+n)$-brane ($3+n+1$ dimensional
manifold) embedded in a $(4+D)$ dimensional bulk \cite{NPB550}. Since
$n$ spatial dimensions are
compact, in this framework, effective 4-dimensional theory also
contain the Kaluza-Klein (KK) excitations of SM fields. The
phenomenology of these models are extensively studied in
Ref. \cite{PRD66}. The volume of $n$ spatial dimensions (internal to
the bulk) can not be too large due to the experimental lower bound on the
KK-mode masses. There are also models in which the
SM particles are confined to a $3$-brane which is ``fat'' i.e. it
has an extension in the $(4+D)$ dimensional bulk \cite{PLB482}.   

On the other hand, there are
class of models where some or all of the SM fields can access
the full space-time manifold.  One such example is Universal Extra
Dimension (UED), where all the fields can propagate in the full
manifold \cite{UED}. Apart from the rich phenomenology, UED models in
general offer possible unification of the gauge couplings at a
relatively low scale of energy, not far beyond the reach of the next
generation colliders \cite{unificUED}.  Moreover, particle spectra of
UED models naturally contain a weakly interacting stable massive
particle, which can be a good candidate for cold dark matter
\cite{dark_ued5,dark_ued6}. 

A particular variant of the UED model where all the SM fields
propagate in $(5 + 1)$ dimensional space-time, namely the {\em two
Universal Extra Dimension} (2UED) model has some attractive
features. 2UED model can naturally explain the long life time for
proton decay \cite{dobrescu} and more interestingly it predicts that the
number of fermion generations should be an integral multiple of three
\cite{dobrescu1}. When the $(5+1)$ dimensional theory is compactified
into a $(3+1)$ dimensional effective theory, each 6D field decomposes
to a tower (known as KK-tower) of 4D fields. Each field
in the KK-tower is characterized by a pair of integers $(j,k)$, known
as KK-numbers. The zero mode (with $j=k=0$) fields are identified with
the SM particles. In this article we will concentrate on the
phenomenology of the $(1,0)$-mode (the lightest KK-mode) particles.

Recently, signals of 2UED model in future colliders
like LHC \cite{dobrescu4, dobrescu3, KGAD1} and ILC \cite{kong, KGAD2}
have been studied in some details. The pair production of strongly
interacting $(1,0)$ modes at the LHC was previously discussed in
Ref. \cite{dobrescu4}. The production of strongly interacting $(1,0)$
modes results into {\em multi lepton + multi jet + missing transverse
  momentum} signal. It is needless to mention that production rate for
strongly interacting particles are high. However, detecting {\em
  multi lepton + missing
  transverse momentum} signal in presence of more than one jets could be
challenging. At the same time it is also important to look for other
smoking gun signature of
this model in other channels and
correlating it with the signal from strong
production channel. Therefore, in this article we
concentrate on the hadronically quiet\footnote{In an environment like
  LHC, even a hadronically quiet process, like the one of our interest, is always associated with one or more soft jets due to initial state radiation. Thus in such an environment, one should ask for the absence of any hard jets (say
  within $p_T>30$ GeV). An event generator like PYTHIA can easily generate such events with soft jets even the hard partonic process does not have any partons in the final state. In our analysis which is done only
  at partonic level, we could not implement this.} signals which
can arise from the 
pair production of the $(1,0)$-mode electroweak particles. We show that the 
pair production of weakly interacting $(1,0)$-modes of 2UED at the LHC
gives rise to {\em multi lepton + missing transverse momentum}
($p_T\!\!\!\!\!/~$) signal. We
concentrate only on {\em 2 (3)-lepton + $p_T\!\!\!\!\!/~~$}
signal. The hypercharge gauge boson $\bmu^{(1,0)}$ decays to
$\bh^{(1,0)}$ and a {\em photon}. The production of
$\bmu^{(1,0)}$ in association with $\bh^{(1,0)}$ can give rise to
$\gamma E\!|!\!\!/~~$ signal at the LHC. This is a very
characteristic feature of 2UED. However, $\bmu^{(1,0)}$ and
$\bh^{(1,0)}$ coupling to quarks, being proportional to the quark
hypercharge, $\bh^{(1,0)} \bmu^{(1,0)}$ production rate is suppressed
at the LHC. Consequently, we
investigate $\gamma~+~E\!\!\!\!/~~$ signal from
$\bh^{(1,0)}\bmu^{(1,0)}$ production in the context of future
$e^{+}e^{-}$ collider. 

The plan of the article is the following. We will give a brief
description of the model in
the next section. Signature of $(1,0)$-mode electroweak particles at
the LHC and ILC will be discussed in section 3 and section 4
respectively. We summarize in the last section.

\section{Two Universal Extra Dimensions}

In 2UED all the SM fields can propagate
universally in the $(5+1)$ dimensional space-time. Four space time
dimensions with coordinates $x^{\mu}$ ($\mu=0,1,2,3$) form the
Minkowski space. Two extra spacial dimensions ($x^4$ and $x^5$) are
flat and are compactified with  
$0\le~x^4,~x^5~\le L$. This implies that the extra dimensional
 space (before compactification) is a square\footnote{This implies
   that the size of the two 
  extra dimensions are same. 
   However, the most general 2UED model should include two different
   sizes for two compactified dimensions instead of one. In
   the absence of any obvious symmetry
 that can relate these two length-scales, we
 are thus considering only a specific choice of parameters of this theory.} with sides $L$. 
Identifying the opposite sides of the
square will make the compactified manifold a torus. However, toroidal
compactification, leads to 4D fermions that are vector-like with
respect to any gauge symmetry. The alternative is to identify 
two pairs of adjacent sides of the square:

\begin{equation}
 (y,0)~\equiv~(0,y),~~~(y,L)~\equiv~(L,y),~~~\forall~y\in~[0,L]
\label{b_cond}
\end{equation}
This is equivalent to folding the square along a
diagonal and gluing the boundaries. Above compactification mechanism
automatically leaves at most a single 4D fermion of definite chirality
as the zero mode of any chiral 6D fermion \cite{dobrescu2}. The physics at identified points is
identical if the Lagrangian takes the same value for any field
configuration:
\begin{eqnarray}
{\cal L}\vert_{x^\mu,y,0}\;=\;{\cal L}\vert_{x^\mu,0,y};~~
{\cal L}\vert_{x^\mu,y,L}\,=\,{\cal L}\vert_{x^\mu,L,y} \nonumber
\end{eqnarray}
This requirement fixes the boundary conditions for 6D scalar fields
 $\Phi(x^\alpha)$ and 6D Weyl fermions $\Psi_{\pm}(x^\alpha)$. The
 requirement that the boundary conditions for 6D scalar or fermionic
 fields are compatible with the gauge symmetry, also fixes the boundary
 conditions for 6D gauge fields. The folding boundary conditions do
 not depend on continuous parameters, rather there are only eight
 self-consistent choices out of which one particular choice leads to
 zero mode fermions after compactification. Any 6D field (fermion/gauge
 or scalar) $\Phi(x^\mu,x^4,x^5)$ can be decomposed as:

\begin{equation}
 \Phi(x^\mu,x^4,x^5)~=~\frac{1}{L}\sum_{j,k}f^{(j,k)}_n(x^4,x^5)
\Phi^{(j,k)}(x^\mu)
\label{kk_dcomp}
\end{equation}
Where,
\begin{equation}
 f^{(j,k)}_n(x^4,x^5)~=~\frac{1}{1~+~\delta_{j,0}\delta_{k,0}}\left[e^{-in\pi/2}cos\left(\frac{jx^4+kx^5}{R}~+~\frac{n\pi}{2}\right)~+~cos\left(\frac{kx^4-jx^5}{R}~+~\frac{n\pi}{2}\right)\right]
\label{kk_func}
\end{equation}
The compactification radius $R$ is related to the size, $L$, of the 
compactified space as : $L = \pi R$. Where 4D fields
$\Phi^{(j,k)}(x^\mu)$ are the $(j,k)$-th KK-modes of the 6D field
$\Phi(x^{\alpha})$ and $n$ is a integer 
whose value is restricted to $0,1,2$ or $3$ by the boundary
conditions. Since $f^{(j,k)}_n(x^4,x^5)$ should form a complete set on the
compactified manifold, it must satisfy the following:
\be
\frac{1}{L^2}\sum_{j,k}\left[f_{n}^{(j,k)}(x^4,x^5)
  \right]^*f_{n}^{(j,k)}(x^{\prime4},x^{\prime 5}) ~=~\delta(x^{\prime
4} - x^{4}) \delta(x^{\prime 5} - x^{5}) 
\label{comp}
\end{equation}
The allowed values of $j$ and $k$ should be chosen such that the
completeness condition in Eq. \ref{comp} is satisfied. It is clear
from the form of $f_{n}^{(j,k)}$ that the functions $f_{n}^{(1,0)}$
and $f_{n}^{(0,1)}$ are not independent ($f_{n}^{(0,1)}=(-1)^n
f_{n}^{(1,0)}$). Therefore, it is sufficient to take $j > 0,~k \ge 0$
and $j=k=0$ to form a complete set of functions on the chiral square. 
It is also obvious from the form of $f^{(j,k)}_n(x^4,x^5)$ that
only $n=0$ allows zero mode ($j=k=0$) fields in the 4D effective
theory. The zero mode fields and the interactions among zero modes can
be identified with the SM.

In 6D, the Clifford algebra is generated by six anticommuting
matrices: $\Gamma^{\alpha},~ \alpha=0,1,..,5$. The minimum
dimensionality of $\Gamma$ matrices in 6D is $8\times8$. The spinor
representation of the $SO(1,5)$ Lorentz symmetry is reducible and
contain two irreducible Weyl representation characterized by different
eigenvalues of the 6D chirality operator: $\bar \Gamma =
\Gamma^0\Gamma^1\Gamma^2\Gamma^3\Gamma^4\Gamma^5$. The chirality
projection operator is defined as: $P_{\pm}=(1\pm\bar\Gamma)/2$, where
$+$ and $-$ label the 6D chiralities defined by the eigenvalues of
$\bar \Gamma$. 
\begin{equation}
\Psi_{\pm}(x^{\alpha})= P_{\pm}\Psi(x^\alpha);~~~~~~~~ \bar \Gamma \Psi_{\pm}
(x^\alpha) = \pm \Psi_{\pm}(x^{\alpha}).
\label{chirality}
\end{equation}
The chiral fermions in 6D have four components. Each 6D chiral fermion
contains both the chiralities of $SO(1,3)$.

Now we move on to the Standard Model in 6-dimensions. In 6D, the fields
and boundary conditions are chosen such that upon compactification,
the zero modes of the resulting effective theory should reproduce the
SM. The requirements of anomaly cancellation and fermion mass generation
 force the weak-doublet fermions to have opposite {\em 6D chiralities}
 with respect to the weak-singlet fermions. So the quarks of one
 generation are given by $Q_+~\equiv~(U_+,D_+),~U_-,~D_-$. Since
 observed quarks and leptons have definite 4D chirality, an immediate
 constraint is imposed on the boundary conditions of doublet and
 singlet fermions. 
 The 6D doublet quarks and leptons decompose into a
 tower of heavy vector-like 4D fermion doublets with left-handed zero
 mode doublets.  Similarly each 6D singlet quark and lepton decompose
 into the towers of heavy 4D vector-like singlet fermions along with
 zero mode right-handed singlets. These zero mode fields are
 identified with the SM fermions. As for example, SM doublet and
 singlets of 1st generation quarks are given by
 $(u_{L},d_{L})~\equiv~Q_{+L}^{(0,0)}(x^{\mu})$, $u_{R}~\equiv~U_{-R}^{(0,0)}(x^{\mu})$
 and $d_{R}~\equiv~D_{-R}^{(0,0)}(x^{\mu})$.

 In 6D, each of the gauge fields, has six
 components. Upon compactification, they decompose into  towers of 4D
 spin-1 fields, a tower of spin-0 fields which are eaten by heavy
 spin-1 fields. Another tower of 4D spin-0 fields, all belonging to
 the adjoint representation of the corresponding gauge group, remain
 in the physical spectrum. These are the physical {\em spinless
 adjoints}. The 6D gluon fields, $G_{\alpha}^{a}$ decompose into a tower
 of 4D spin-1 fields, $G_{\mu}^{a(j,k)}$, and a tower of spin-0 fields,
 $G_{H}^{a(j,k)}$. $G_{\mu}^{a(j,k)}$ tower includes a zero mode which
 can be identified with the SM gluons. Similarly 6D $SU(2)$ gauge
 fields have KK-modes $W_{\mu}^{(j,k)\pm}$, $W_{H}^{(j,k)\pm}$,
 $W_{\mu}^{(j,k)3}$ and $W_{H}^{(j,k)3}$, while the hypercharge gauge
 field has KK-mode $B_{\mu}^{(j,k)}$ and $B_{H}^{(j,k)}$. The zero
 modes of $W_{\mu}^{(j,k)\pm}$ towers are identified with the SM
 $W^{\pm}_{\mu}$ bosons. The mixing of $W_{\mu}^{(0,0)3}$ and
 $B_{\mu}^{(0,0)}$ gives photon and $Z$-boson. However, for non-zero
 modes this mixing is negligible.

The tree-level masses for $(j,k)$-th KK-mode particles are given by
$\sqrt{M_{j,k}^2~+~m_{0}^{2}}$, where $M_{j,k}=\sqrt{j^2+k^2}/R$.
 $m_0$ is the mass of the corresponding zero mode particle. As a result,
 the tree-level masses are approximately degenerate. This degeneracy
 is lifted by radiative effects. 
 The fermions receive mass corrections from the gauge interactions 
 (with gauge bosons and adjoint scalars) and Yukawa interactions. 
All of these give positive mass shift. The gauge fields and spinless 
adjoints receive mass corrections from the self-interactions and gauge
 interactions. Gauge interactions with fermions give a negative mass 
shift. While the self-interactions give positive mass shift with 
different strength with respect to the former. However, masses of the 
hypercharge gauge boson $B_{\mu}^{(j,k)}$ and the corresponding 
scalar $B_{H}^{(j,k)}$ 
receive only negative corrections from fermionic loops. Numerical 
computation shows that the lightest KK particle is the spinless adjoint 
$B_{H}^{(1,0)}$, associated with the hypercharge gauge boson. As a result, 
2UED model gives rise to a scalar dark matter.

\subsection{$(1,0)$-mode electroweak sector of 2UED}
 The $(1,0)$-mode electroweak sector of 2UED consists of
 $(1,0)$-mode gauge bosons ($\bmu^{(1,0)},~W_{\mu}^{3(1,0)}$ and
 $~W_{\mu}^{\pm(1,0)}$), spinless adjoints
 ($\bh^{(1,0)},~W_{H}^{3(1,0)}$ and $~W_{H}^{\pm(1,0)}$) of $U(1)$ and
 $SU(2)$ gauge group respectively, $(1,0)$-mode KK excitations of
 the SM leptons and $(1,0)$-mode excitations of the Higgs doublet.
 In the previous section we have qualitatively discussed the effects
 of radiative corrections on the mass spectrum. After incorporating
 the radiative effects, approximate expressions for the masses at the
 $(1,0)$ level can be
 written as 
\bea  
M_{L_{+}}&\simeq&{1.04}~{R^{-1}}, \hskip 100pt
M_{E_{-}}~\simeq~{R^{-1}}, 
\nonumber\\ 
M_{B_{\mu}}&\simeq&{0.97}~{R^{-1}}, \hskip 100pt
M_{B_{H}}~\simeq~{0.86}~{R^{-1}}, 
\nonumber\\ 
M_{W_{\mu}}&\simeq&{1.07}~{R^{-1}}, \hskip 100pt
M_{W_{H}}~\simeq~{0.92}{R^{-1}}.
\label{mass}
\eea
The numerical factors in the above expressions are almost independent
of $R^{-1}$ for 
$(1,0)$-mode leptons, gauge boson and spinless adjoint
corresponding to $U(1)$ gauge group. However, masses of $SU(2)$ gauge
bosons and spinless adjoints do have mild dependencies on $R^{-1}$. Detailed
expressions for the one-loop corrected masses of KK particles in
2UED can be found in Ref. \cite{loop}.

Decays of $(1,0)$-mode particles of 2UED have been previously
investigated in details in Ref. \cite{dobrescu4}. Conservation
of KK-parity allows $(1,0)$-mode particles to decay only into
a $(1,0)$-mode particle and one or more SM particles if kinematically
allowed. It is clear from Eq. (\ref{mass}) that $B_{H}^{(1,0)}$ is the
lightest KK particle (LKP) in this theory. It is important to notice
that, unlike in the case of 1UED, the LKP is a scalar in this
scenario. Since $\bh^{(1,0)}$ is a stable particle and weakly
interacting, it passes through the detector without being
detected. Decays of all the $(1,0)$-mode particles thus result into
one or more SM particles plus missing energy/momentum signature.


\begin{table}[h]

\begin{center}

\begin{tabular}{|c|c|c|c|c|c|c|}
\hline \hline
$R^{-1}$ in GeV              & 300 & 350 & 400 & 450 & 480 & 500 
\\\hline \hline 
$B_{\mu} \to l \bar l B_{H}$ & 0.610 & 0.618 & 0.624 & 0.630 & 0.633 & 0.635
\\\hline 
$B_{\mu} \to \gamma B_{H}$   & 0.390 & 0.382 & 0.375 & 0.369 & 0.364 & 0.362
\\\hline\hline

\end{tabular}

\end{center}

\caption{Branching fractions of $B_{\mu}$ in $l \bar l B_{H}$ and
  $\gamma B_{H}$ channel for different values of $R^{-1}$} 

\label{bmu_br}
\end{table}

            
Let us
begin with the $U(1)$ gauge boson $\bmu$\footnote{From now on we will
  only concentrate on $(1,0)$-mode, thus drop the $(1,0)$
  superscript.}. $\bmu$ dominantly decays to
two SM charged leptons and $\bh$. This is a tree level 3-body
decay. Apart from this 3-body decay, loop induced decay into a
photon and $\bh$ has comparable branching fraction
\cite{dobrescu4}. $\bmu$ decay to $f\bar f \bh$ is 3-body decay
mediated by corresponding $(1,0)$ mode fermion. Decay amplitude is
proportional to the hypercharge of the fermions in consideration. As a
result, $\bmu~\to ~\nu \bar \nu \bh$ and $\bmu~\to ~q \bar q \bh$ are
suppressed compared to $\bmu~\to ~l \bar l \bh$. This is even
applicable for $\bmu~\to ~u_R \bar u_R \bh$ decay (hypercharge of
$u_R$ is $4/3$). This is accounted by the fact that in case of
$\bmu~\to ~u \bar u \bh$, decay amplitude is suppressed by heavier
$U^{(1,0)}$ propagator (than $L^{(1,0)}$ propagator in case of
$\bmu~\to ~l \bar l \bh$). In fact decay amplitude is inversely
proportional to the mass difference of $\bmu$ and the propagator mass.
In Table 1, we have tabulated the branching fractions of $\bmu$ in $l 
\bar l \bh$ (where $l$ includes $e,~\mu$ and $\tau$) and $\gamma \bh$
channel for different values of $R^{-1}$. $SU(2)$ spinless
adjoints ($W^{\pm}_H,~W^3_H$) can decay only to the $B_{H}$ and SM
particles. $W^3_H$
decays to a pair of SM leptons and $\bh$ with equal branching ratio to
charged leptons and neutrinos. Branching 
fraction to quark antiquark pairs is again negligible due to
hypercharge and heavy $(1,0)$ mode quark propagator. $W^{\pm}_{H}$
decay with almost 100\% branching ratio to $ l \bar\nu_{l} \bh$ ($l$
includes all 3 SM lepton generations). Branching fractions of $SU(2)$
spinless adjoints are independent of $R^{-1}$.  

\begin{table}[h]

\begin{center}

\begin{tabular}{|c|c|c|c|c|}
\hline \hline
Particle & \multicolumn{4}{c|}{Branching fractions in } \\\cline{2-5}
         &  $0l~+~E\!\!\!/$ & $1l~+~E\!\!\!/$ &
$2l~+~E\!\!\!/$ & $3l~+~E\!\!\!/$\\\hline\hline
$W^{3}_{\mu}$  & 0.16 & 0 & 0.72 & 0 \\
$W^{\pm}_{\mu}$  & 0 & 0.55 & 0 & 0.42\\\hline\hline

\end{tabular}

\end{center}

\caption{Branching fractions of $(1,0)$-mode $SU(2)$ gauge bosons in
  {\em multi lepton + missing energy ($E\!\!\!/~$)} channel for
  $R^{-1}=500$ GeV}

\label{wmu_br}
\end{table}

Since the $(1,0)$-mode $SU(2)$ gauge bosons ($W^\pm_\mu,~W^3_\mu$) are
heavier than $(1,0)$-mode 
leptons (see Eq. \ref{mass}), they decay dominantly into
$(1,0)$-mode lepton doublets and corresponding SM leptons
\cite{dobrescu4}. As for
example, $W^3_\mu$ can decay into one of the six ($l_i L_i^{(1,0)}$ and
$\nu_i \nu_i^{(1,0)},~i=e,\mu,\tau$) channels with equal
probability. Similarly, 
$W^\pm_\mu$ decays into one of the six possible decay modes ($l_i
\nu_i^{(1,0)}$ and $\nu_i L_i^{(1,0)},~i=e,\mu,\tau$) with branching
fraction of $1/6$ into each decay modes. The $(1,0)$-mode leptons are
heavier than  
$\bmu$, $\bh$ and $SU(2)$ spinless adjoints. Therefore,
$L^{(1,0)},~\nu^{(1,0)}$ can decay 
into the corresponding SM lepton and $\bmu~ (\bh)$ or $SU(2)$ spinless
adjoints. Following
this decay chain, one can see that $(1,0)$-mode $SU(2)$ gauge bosons
dominantly decay to
$\bh$ and one or more SM leptons. We have presented the branching
fractions of $(1,0)$-mode $SU(2)$ gauge 
bosons in multi lepton plus missing energy channels for
$R^{-1}=500~$GeV in Table \ref{wmu_br}.
Branching ratios of $SU(2)$ gauge bosons are almost independent over
$R^{-1}$. As for example, branching fraction of $W_{\mu}^{3}\to~2l~+
E\!\!\!/$ increases by 0.11\% when $R^{-1}$
is changed from 300 to 500 GeV.

One may tempted to think that the existing bound of $Z^\prime$-mass
(present in an extension of the SM with an extra $U(1)$ symmetry) might
be applicable also to $\bmu~(W_{\mu}^3)$. For example,
$Z^\prime$ can be produced at resonance at the Tevatron and can decay
to $e^+e^-$ or $\mu^+\mu^-$ pairs \cite{PDG}. The present bound on
$Z^\prime$-mass is around $900$ GeV. However, $\bmu~(W_\mu^3)$ can not be
produced singly (due to KK-parity conservation) from $p\bar p$
collisions. At the same time $\bmu~(W_\mu^3)$, once produced, decays
to some SM particles (photon or leptons) and $\bh$. As a result, any
decay of $\bmu~(W_\mu^3)$ is always associated with large missing
momentum/energy. Thus the direct/indirect search limits on $Z^\prime$
are not applicable to $\bmu~(W^3_\mu)$.  

\section{Signature of electroweak $(1,0)$-mode particles at the LHC}

In this section, we will first discuss the production of electroweak
$(1,0)$-mode particles of 2UED at the LHC. Phenomenology in 2UED is
different and
perhaps more complicated compared to 1UED due to the presence of spinless
adjoints. In a previous work \cite{KGAD1} we have studied
the production (in the context of LHC) and decays of $(1,1)$-mode
spinless adjoints in some details. 

\begin{figure}[t]
\begin{center}
\epsfig{file=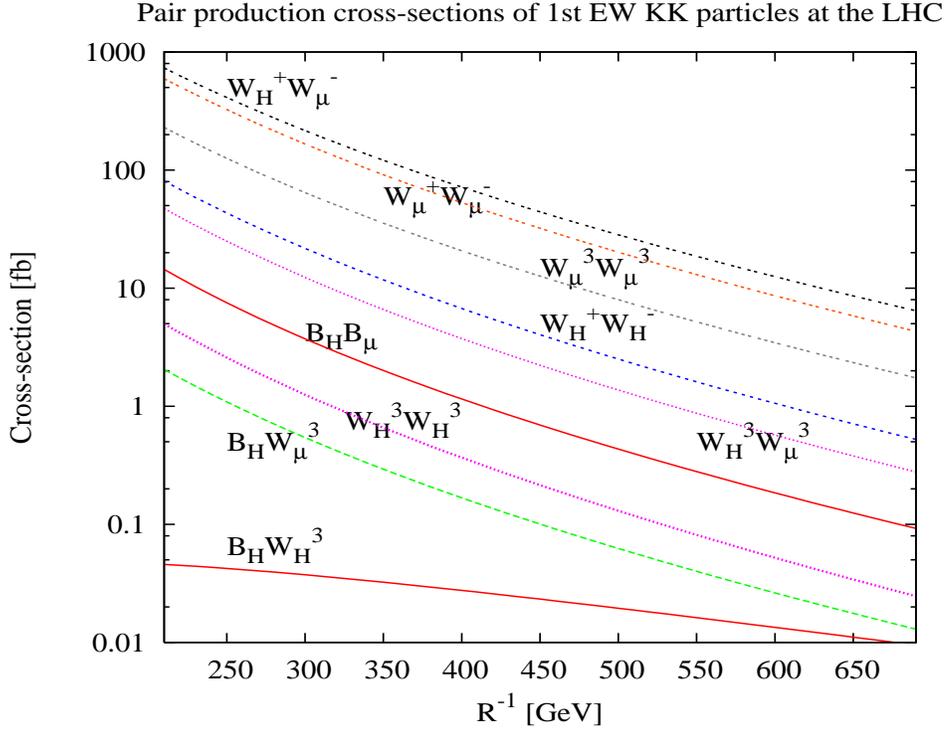,width=10cm,height=13cm,angle=270}
\end{center}
\caption{Pair production cross-sections of $(1,0)$-mode
  states
at the LHC as a function of
  $R^{-1}$. CTEQ4L parton distribution functions are used to evaluate
  those cross-sections.} 
\label{cross_LHC}
\end{figure}

Due to the conservation of KK-parity, single
production of $(1,0)$-mode particles is not possible. So they must be
produced in pairs. All the $(1,0)$-mode electroweak
gauge bosons and spinless adjoints have tree level couplings with an
$(1,0)$-mode fermion and a SM fermion. These couplings arise
from the compactification of 6D kinetic term for fermions, present in
the 6D bulk Lagrangian. Two $(1,0)$-mode charged $SU(2)$ gauge
bosons or spinless adjoints can couple to a SM photon or a SM
$Z$-boson. Those couplings arise from the compactification of the
kinetic term for 6D non Abelian gauge
fields. All the vertices relevant for the pair production of
$(1,0)$-mode electro-weak particles can be found in Appendix A. Pair
production of $(1,0)$-mode electroweak gauge bosons or
spinless adjoints at the LHC take place via the above mentioned
interactions only. We have estimated the production
cross-sections of the following pairs of neutral $(1,0)$-mode states:
$\sigma(\bmu,\bh)$, $\sigma(\wmu,\bh)$, $\sigma(\wh,\bh)$, $
\sigma(\wmu,\wh)$, $\sigma(\wh,\wh)$, $\sigma(\wmu,\wmu)$; pairs of
opposite charged $(1,0)$- modes: $
\sigma(\wmupm$, $W_{\mu}^{\mp})$, $\sigma(\whpm,W_{H}^{\mp})$,
$\sigma(\wmupm,W_{H}^{\mp})$ and pairs of one charged and one neutral
states: 
$\sigma(\wmupm,\wmu)$, $\sigma(\wmupm,\wh)$, 
$\sigma(\whpm,\wmu)$, $\sigma(\whpm,\wh)$ in proton-proton collision
at center-of-mass energy of $14$ TeV. CTEQ4L parton distribution functions
\cite{cteq4} are used to numerically evaluate the above
cross-sections. The factorization scale (for parton distribution
functions) is fixed at the $(1,0)$-mode mass. For this study we have
used one-loop corrected mass spectrum for $(1,0)$-mode particles given
in Eq. (\ref{mass}). 

\begin{figure}[t]
\begin{center}
\epsfig{file=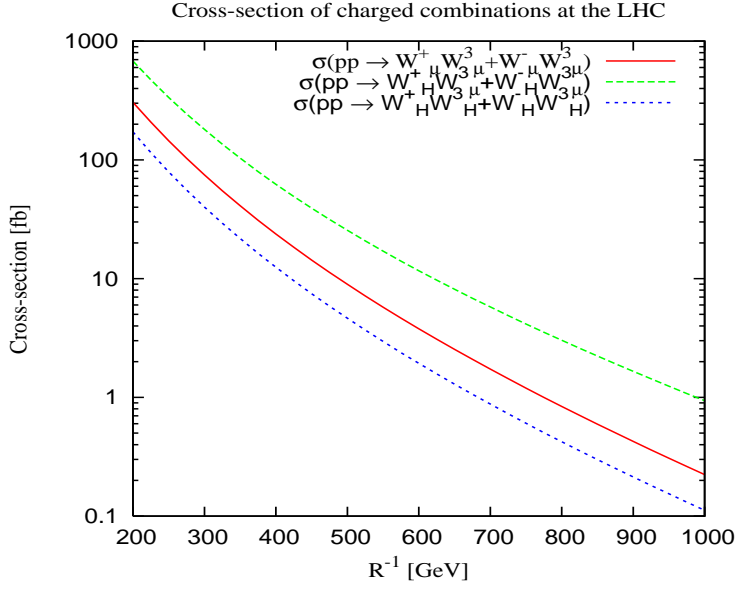,width=8cm,height=10cm,angle=270}
\end{center}
\caption{Production cross-sections for one charged $+$ one neutral
  $(1,0)$-modes 
at the LHC as a function of
$R^{-1}$. CTEQ4L parton distribution functions are used to evaluate
those cross-sections.} 
\label{cross_charged}
\end{figure}

The production cross-sections of two neutral $(1,0)$-mode or two
oppositely charged $(1,0)$-mode final particles are
presented in
Fig. \ref{cross_LHC} as function of $R^{-1}$. The processes presented in
Fig. \ref{cross_LHC}, apart from being completely driven by
electroweak couplings, are initiated by a quark and an antiquark. LHC,
being a proton-proton collider, antiquarks can only arise from
sea-excitations. Consequently, at the energy scale of our interest, the density of
antiquarks inside a proton is small compared to quarks. This makes
the above cross-sections small. 

Production cross-sections of $U(1)$ spinless adjoint ($\bh$) in
association with $SU(2)$ spinless adjoint ($\wh$) or gauge boson
($\wmu$) are smallest among the others. $\bh\wh~(\wmu)$ pair
production cross-section vary from 0.05 (2.5) fb to 0.01 fb as as we
vary $R^{-1}$ from 200 to 650 (700) GeV. $\sigma(\wh\wh)$ and
$\sigma(\bmu\bh)$ are
large compared to the previous two but the numerical values are not
very promising. $\wh\wh~(\bmu\bmu)$ production cross-section varies
from 8.75 (14.4) fb to 0.06 (0.09) fb as we vary $R^{-1}$ from 200 to
690 GeV. At the parton level, $q \bar q\to\bh \wh ~(\wmu)$ or $\wh\wh$
process is mediated by an excited $(1,0)$-mode quark in the $t~
(u)$-channel. Since 
$SU(2)$ gauge fields in 4D (6D) couple only with left-handed (6D +ve
chirality) fermions, only the +ve chirality $(1,0)$-mode fermion
$Q^{(1,0)}_{+}$ contributes in $t~(u)$-channel. The coupling of a
$(1,0)$-mode $U(1)$ gauge boson or spinless adjoint with an
$(1,0)$-mode quark and a SM quark is proportional to the hypercharge
of the corresponding quark. Due to small hypercharge of the quarks,
processes involving one or two $U(1)$ gauge boson or spinless adjoint
are suppressed at hadron collider.

Rest of the cross-sections, shown in Fig. \ref{cross_LHC}, involves
$SU(2)$ gauge bosons and spinless adjoints. These cross-sections are
large even for larger value of $R^{-1}$. As for example, cross-section
for $\wmupm W_{H}^{\mp}~(W_{\mu}^{\mp})$ production and $\wmu$ pair
production varies from few hundred femtobarn to few femtobarn as we
vary $R^{-1}$ from 200 to 700 GeV.

The production cross-sections of one charged and one neutral
  $(1,0)$-modes are presented
in Fig. \ref{cross_charged}. $\wmupm\wh$ and
$\whpm\wmu$ production cross-sections vary
from few hundred femtobarn to $1$ fb as we vary $R^{-1}$ from $200$
GeV to $1$ TeV. In Fig. \ref{cross_charged} we have not presented the
production cross-sections of charged $SU(2)$ gauge bosons or spinless
adjoints in association with a $U(1)$ gauge boson or spinless
adjoint. Those production cross-sections are very much suppressed
compared to the others due to the small hypercharge of quarks.  


\begin{table}[h]

\begin{center}

\begin{tabular}{|c|c|}
\hline \hline
Final State & Parton level sub-processes \\
            &  $q\bar q \to AB$\\\hline\hline
 & $\bmu \bh$, $\wmu \bh$, $\wh \bh$,\\
$2-lepton~+~p_T\!\!\!\!\!/~~$ &$\wmu \wh$, $\wh \wh$, $\wmu \wmu$,\\
& $\wmupm W_{\mu}^{\mp}$, $\whpm W_{H}^{\mp}$, $\wmupm
W_{H}^{\mp}$\\\hline 
$3-lepton~+~p_T\!\!\!\!\!/~~$ & $\wmupm \wmu$, $\wmupm\wh$,
$\whpm,\wmu$,\\ 
& $\whpm,\wh$ \\\hline\hline

\end{tabular}

\end{center}

\caption{List of parton level sub-processes that contribute to the
  $2-lepton~+~p_T\!\!\!\!\!/~~$ and $3-lepton~+~p_T\!\!\!\!\!/~~$
  signal at the LHC.} 

\label{list}
\end{table}


After a very short discussion about the pair production
cross-sections of
$(1,0)$-mode electroweak bosons and spinless-adjoints, we will now
analyze the possible 
signals of this sector at the LHC. Electroweak
$(1,0)$-mode particles of 2UED exclusively decay to multi
lepton and $\bh$. Only exception is the $U(1)$ gauge boson
$\bmu$, which can decay into a photon and $\bh$. Therefore,
the pair production of $\bmu$ and production of $\bmu$
in association with $\bh$ give rise to two photon and one
photon +  $p_{T}\!\!\!\!\!\!/~~$ signal respectively. One or two
photon +
$p_{T}\!\!\!\!\!\!/~~$ signals are unique for 2UED. However, due to small
hypercharge of quarks, these production cross-sections are small
at the LHC. Thus it is not possible to detect photon plus
$p_{T}\!\!\!\!\!\!/~$ signal over the SM background. Since the
hypercharges of electron and positron are large, the production
cross-sections of $\bmu$ in association with a $\bh$ is expected to be
large at an $e^+e^-$ collider. Therefore, $\bmu \bh$ pair production
may gives rise to interesting signals of 2UED at $e^+e^-$ collider. We
will consider this possibility in Section 4 in details.

Since all other electroweak $(1,0)$-mode spinless adjoints or gauge
bosons almost exclusively decay to SM leptons and $\bh$, we
concentrate on {\em two SM leptons + $p_{T}\!\!\!\!\!\!/~~$} signal
resulting from the production of two neutral or two oppositely charged
$(1,0)$-mode particles and {\em
  three SM lepton + $p_{T}\!\!\!\!\!\!/~~$} signal arising
from the production of one charged plus one neutral $(1,0)$-mode
states. In Table \ref{list}, we have listed all the parton level
sub-processes that contribute (after decay) to the {\em two
  (three) SM leptons + $p_{T}\!\!\!\!\!\!/~~$} signal. The decays of
electroweak $(1,0)$-modes are discussed in section 2.1 (see Table
\ref{bmu_br} and Table \ref{wmu_br}).

\subsection{2-lepton + $p_{T}\!\!\!\!\!\!/~~$ signal}

2-lepton + $p_{T}\!\!\!\!\!\!/~~$ signal arises from the production of
two (both neutral or charged) $(1,0)$ mode electroweak particles
(presented in Fig. \ref{cross_LHC}).
Since the tau lepton
detection efficiency 
is significantly different from both electron and muon, the signal
is summed over electron and muon only ($l=e,~\mu$).

\begin{table}[h]

\begin{center}

\begin{tabular}{c|c|c}
\hline\hline

Kinematic Variable & Minimum value & Maximum value \\\hline\hline
$\Delta R(l^{+}l^{-})$ & 0.3  & -    \\
$p_{T}^{l^{+},l^{-}}$  & 10 GeV   & -    \\
$\eta_{l^{+},l^{-}}$ & -2.5  & 2.5 \\
$p_{T}\!\!\!\!\!\!/~$  & 25 GeV   & -  \\
$M(l^{+}l^{-})$    & 10 GeV      & - \\\hline\hline

\end{tabular}

\end{center}

\caption{Acceptance cuts on the kinematical variables for {\em
2-lepton + $p_T\!\!\!\!\!\!/~~$} signal.}

\label{cut_acc}
\end{table}


We have used a parton level Monte-Carlo computer code to evaluate the
$2l+p_T\!\!\!\!\!/~~$ and $3l+p_T\!\!\!\!\!/~~$ cross-sections. To
parametrize detector acceptance and enhance signal to background ratio
, we have 
imposed kinematic cuts, listed in Table \ref{cut_acc}. It is important
to mention that production of 
$(1,0)$-mode lepton pairs and their decay to a SM lepton + $\bh$ also
contributes to the signal. However, production of $(1,0)$-mode lepton
pairs at a hadron collider is purely a s-channel process and thus is
very much suppressed.  

With the acceptance cuts, defined in Table \ref{cut_acc}, the total
signal cross-section is small 
for the higher values of $R^{-1}$. As for example, the total signal
cross-section is $19.46$ fb for $R^{-1}=450$ GeV.
However, the difficulty in detecting the
signals is not the small rate of
production, but due to the large SM background which we will discuss
in the following.

In the SM, the dominant contribution to $2l+ p_{T}\!\!\!\!\!\!/~~$
comes from the $W$-boson pair production, $Z$-boson pair production
and production of a Z-boson in association with a virtual
photon. $2l+ p_{T}\!\!\!\!\!\!/~~$ background can also come
from the production of a $WZ$-pairs followed by leptonic decay of both
$W$ and $Z$ where one of the charged lepton falls outside the detector
coverage. However, with the acceptance cuts (listed in Table
\ref{cut_acc}) this cross-section is estimated to be very small.

\begin{table}[h]

\begin{center}

\begin{tabular}{c|c|c}
\hline\hline

Kinematic Variable & Minimum value & Maximum value \\\hline\hline
$M(l^{+}l^{-})$    & 10 GeV      & 85 GeV \\
$\Delta \Phi_{l^+ l^-}$ &   -    & $177^0$\\\hline\hline

\end{tabular}

\end{center}

\caption{Selection cuts on the kinematical variables for {\em
2-lepton + $p_T\!\!\!\!\!\!/~~$} signal.}

\label{sec_2l}
\end{table}


\begin{figure}[t]

\begin{center}
\epsfig{file=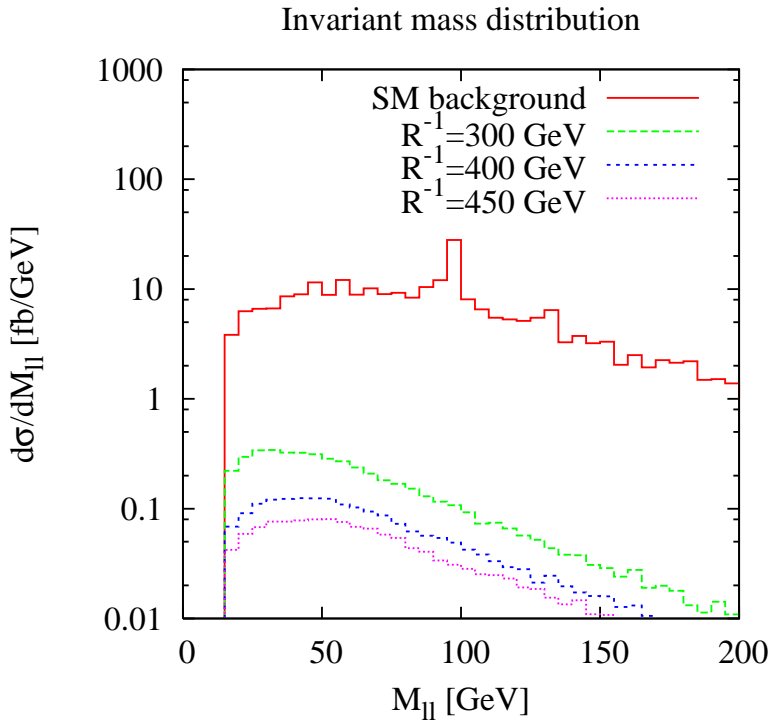,width=10cm,angle=270}
\end{center}
\caption{Invariant mass (of lepton pair) distribution of signal
  (for three different values of $R^{-1}$) and background (for {\em
2-lepton plus $p_T\!\!\!\!\!\!/~~$} signal).}
\label{inv_mas}

\end{figure}

It is important to notice that the signal, we are considering, consists
of only two observable charged leptons which may come from
ordinary Drell-Yan process. However, Drell-Yan production of lepton pairs
is not accompanied by missing energy. So this kind of background can
be removed simply by demanding a minimum value of the missing transverse
momentum. Drell-Yan production of $\tau$-lepton pairs and subsequent
leptonic decays of $\tau$ provides a real background to 2UED
signal. However, the leptons ($l=e,~\mu$) resulting from the
Drell-Yan production of $\tau$ pairs are almost back to back in
the transverse plane. Therefore, this contribution can be
completely removed by demanding an upper bound on the angle between
leptons in the transverse plane.

\begin{table}[t]

\begin{center}

\begin{tabular}{|c|c|c|}
\hline\hline

Contributing & \multicolumn{2}{|c|}{Cross-section in fb} \\\cline{2-3}
SM process   & After acceptance cuts & After selection cuts \\\hline\hline
$pp\to \tau \bar \tau$ & 216 & 4.7 \\\hline
$pp\to W^{\pm}Z$       & 2.1 & 0.4  \\\hline
$pp\to t \bar t$       & 22.5 & 8.13  \\\hline
$pp \to W^+W^-+\gamma^* Z$ &1220  & 444 \\\hline\hline

\end{tabular}

\end{center}

\caption{Standard model background cross-sections for {\em
2-lepton + $p_T\!\!\!\!\!\!/~~$} signal.}

\label{red_bac}
\end{table}


The production $t \bar t$ pairs and subsequent semi-leptonic decays of
both the top quarks also contributes to the $2l+ p_{T}\!\!\!\!\!\!/~~$
background where both the partons are either too soft to be identified
as jets or fall outside detector coverage. We made an estimate of this
background by demanding the partons can be identified as jets only if their
$p_T > 30$ GeV\footnote{We keep in mind that in LHC environment
  purely leptonic signal like ours are always come with soft jets. In
  our parton level analysis, we assume that unless the $b$-parton
  coming from $t$-decay have $p_T>30$ GeV, it can not be identified as
jets.}.


\begin{table}[h]

\begin{center}

\begin{tabular}{|c|c|c|}
\hline \hline
$R^{-1}$ & \multicolumn{2}{|c|}{Signal cross-section in fb}
\\\cline{2-3}
in GeV  &  After acceptance cut & After selection cut \\\hline \hline
300     & 24.4 & 18.5 \\
350     & 19.1 & 11.4 \\
400     & 10 & 7.3\\
450     & 6.7 & 4.7\\ \hline\hline
\end{tabular}
\end{center}

\caption{{\em 2-lepton + $p_T\!\!\!\!\!\!/~~$} signal
  cross-sections for different values of $R^{-1}$.}

\label{signal}
\end{table}


The background from $Z$-boson pair production can be eliminated by 
rejecting events in which the invariant mass of the lepton pairs is
close to the mass of $Z$-boson. Fig. \ref{inv_mas} shows the invariant
mass distribution for the SM background and signal. $Z$-pole is
clearly visible in the SM background distribution. To extract signal
from the SM 
background we impose the selection cuts listed in Table
\ref{sec_2l}. The choice of selection cut on the 
invariant mass of the lepton pairs is largely aimed to reduce
the contribution from the $Z$-boson pair production. With those
selection cuts, listed in Table \ref{sec_2l}, the background, arises from
production of $t \bar t$ pairs, is also found to be small. In Table
\ref{red_bac}, we have presented the background
cross-sections after acceptance cuts (second column) and selection
cuts (third column).
    
The dominant
background to the $2l+p_T\!\!\!\!\!\!/~~$ signal is $W$-boson pair
production from 
quark-antiquark annihilation followed by leptonic decay of each of the 
$W$-bosons. It is difficult to choose a selection criteria which can
completely remove
this background without affecting the signal.

Signal cross-section for different values of $R^{-1}$
are presented in Table \ref{signal} after applying the
acceptance cuts (2nd 
column) and selection cuts (3rd column) respectively. The
corresponding background 
cross-sections are also presented in Table \ref{red_bac}. In
order to quantify the ability of extracting signal event, 
$N_S=\sigma_S{\cal L}$, for a given integrated luminosity ${\cal L}$
over the SM background events, $N_B=\sigma_B{\cal L}$, we define the
significance $S=N_S/{\sqrt N_B}$. It is
clear from the numbers in Table \ref{signal} that, with $100~fb^{-1}$
luminosity, more than $5\sigma$ discovery of the signal is possible upto
$R^{-1}~=~ 350$ GeV.

\subsection{3-lepton + $p_{T}\!\!\!\!\!\!/~~$ signal}

In the frame work of 2UED, {\em 3-lepton + $p_T\!\!\!\!\!\!/~~$}
signal results from the production of a neutral $(1,0)$-mode gauge
boson or spinless adjoint in association with a charged $(1,0)$-mode
$SU(2)$ gauge boson or spinless adjoint (presented in
Fig. \ref{cross_charged}). As for example, the
production of a neutral gauge boson or spinless adjoint in association
with a charged $SU(2)$ spinless adjoint and the subsequent decay of the
charged adjoint in {\em 1-lepton + $p_T\!\!\!\!\!\!/~~$} channel
and neutral particle in {\em 2-lepton + $p_T\!\!\!\!\!\!/~~$}
channel give rise to {\em 3-lepton + $p_T\!\!\!\!\!\!/~~$}
signal, 
$
pp\to\whpm \wmu~(\wh)\to(l^{\pm}\bh)(l^+l^-\bh) 
$.
However, the decay of $(1,0)$-mode $SU(2)$ charged gauge bosons into
{\em 1-lepton + $p_T\!\!\!\!\!\!/~~$} channel and {\em 3-lepton
  + $p_T\!\!\!\!\!\!/~~$} channel are equally probable as
can be seen from Table \ref{wmu_br}. Therefore, for $\wmupm\wmu~(\wh)$
production, both the invisible decay and {\em 2-lepton + $\bh$}
decay of $\wmu~(\wh)$ contributes to the signal. Instead of including all 
three SM lepton generations, in this part of the work we consider
only first two 
generations of SM leptons ($l=e,~\mu$). We impose following selection
criteria, listed in Table \ref{cuts_3l}, on the kinematical variables after
ordering the leptons according to their $p_T$ hardness ($p_T^{l_1}\le
p_T^{l_2}\le p_T^{l_3}$).

\begin{table}[t]

\begin{center}

\begin{tabular}{c|c|c}
\hline\hline

Kinematic Variable & Minimum value & Maximum value \\\hline\hline
$\Delta R(l_i l_j)$, $i\ne j$ & 0.3  & -    \\
$p_{T}^{l_i}$  & 10 GeV   & -    \\
$\eta_{l_i}$ & $-$2.5  & 2.5 \\
$p_{T}\!\!\!\!\!\!/~$  & 25 GeV   & -  \\\hline\hline

\end{tabular}

\end{center}

\caption{Acceptance cuts on the kinematical variables for {\em
3-lepton + $p_T\!\!\!\!\!\!/~~$} signal with $i,j=1,2,3$ and
  $l=e,~\mu$.}

\label{cuts_3l}
\end{table}


\begin{table}[h]

\begin{center}

\begin{tabular}{|c||c||c|c|c|c|}
\hline\hline
 &  & \multicolumn{4}{|c|}{Signal} \\
  &Background & \multicolumn{4}{|c|}{cross-section} \\
Applied Cuts & cross-section & \multicolumn{4}{|c|}{in fb} \\\cline{3-6}
  &  in fb & \multicolumn{4}{|c|}{$R^{-1}$ in GeV}
\\\cline{3-6}
  &                & 400 & 500 & 550 & 600 \\\hline\hline

Acceptance cuts & 145 & 8.83 & 4.51 & 3.29 & 2.44 \\\hline
$10~GeV< p_T^{l_1}< 50~GeV$, & & & & & \\
$15~GeV< p_T^{l_2}< 50~GeV$, & 99.7 & 7.2 & 3.71 & 2.74 & 1.98\\
$p_T^{l_3}>20~GeV$ & & & & & \\\hline 
$20~GeV< M_{l_1 l_2} < 70~GeV$, & & & & & \\
$|M_{l_1 l_3}-m_Z|> 10~GeV$,  & 6.39 & 4.8 & 2.46 & 1.72 & 1.23\\
$|M_{l_2 l_3}-m_Z|> 10~GeV$ & & & & &  \\\hline\hline

\end{tabular}

\end{center}

\caption{{\em 3-lepton + $p_T\!\!\!\!\!\!/~~$} signal (for
  different values of $R^{-1}$) and corresponding SM background after
  three sets of successive cuts. $m_Z$ is the mass of $Z$-boson.}

\label{signal_3l}
\end{table}

\begin{figure}[t]

\begin{center}
\epsfig{file=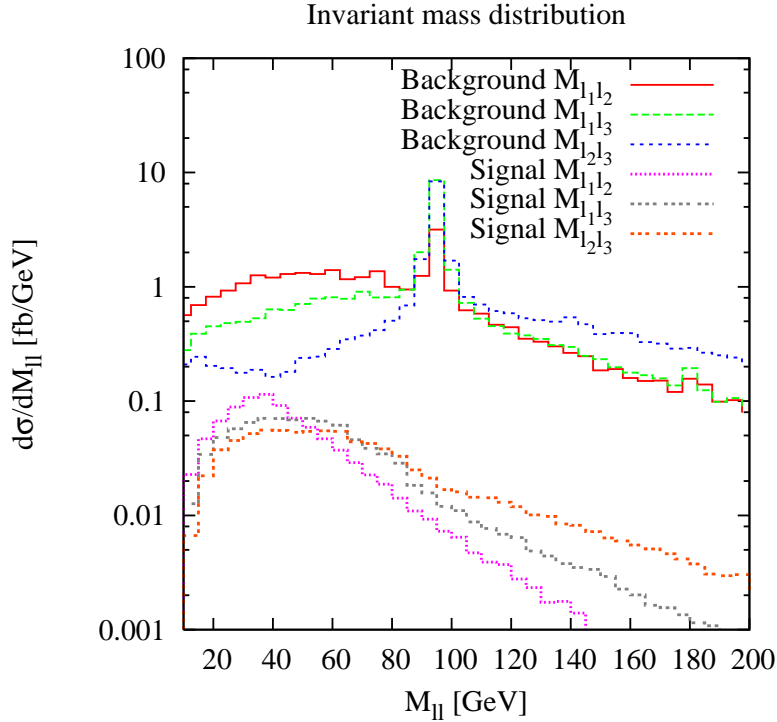,width=10cm,angle=270}
\end{center}
\caption{Invariant mass distribution of signal
(for $R^{-1}=500$ GeV) and background (for {\em
3-lepton + $p_T\!\!\!\!\!\!/~~$} signal).}
\label{inv_mas3l}

\end{figure}

In the SM, the dominant contribution to {\em 3-lepton +
  $p_T\!\!\!\!\!\!/~~$} comes from the production of $W$-boson in
  association with a $Z$-boson or a virtual photon. The $WZ$ production is
  characterized by a peak in the invariant mass distribution of the
  lepton pairs at the $Z$-boson mass. Fig. \ref{inv_mas3l} shows the
  invariant mass (of three possible combinations of leptons)
  distributions for signal (for
  $R^{-1}=500$ GeV) and background. Therefore, the
  contributions from $WZ$ production can be eliminated by rejecting
  the events in which the invariant mass of any lepton pair is close
  to the mass of the $Z$-boson. However, it is difficult to introduce an
  event selection criteria to eliminate the $W\gamma^*$ contribution
  without affecting the signal. However, the lepton pairs coming from
  $\gamma^*$ are in general soft an try to be collinear to each
  other. Hence a cut on the invariant mass of the $l_1 l_2$ can remove
  a large part of this background.

To extract the signal from the SM background we have introduced a set
of cuts, listed in Table \ref{signal_3l}. It is important to notice
that the signal leptons results from the decay of a heavy particle
(such as $W^{\pm}_{\mu},~W_H^{\pm},~W^{3}_{\mu}$ and $W_H^{3}$)
into another heavy particle ($\bh$) plus one or more SM
leptons. Therefore, the SM leptons most of the time carry small amount
of energy. However,
the background leptons arises from the decay of $W$ and $Z$ or a virtual
photon. Consequently, the leptons from the decay of $W$ or $Z$ bosons are not
very much soft. We have exploited this feature of the signal to
enhance the signal to background ratio. In Table \ref{signal_3l}, we have
imposed some upper bounds on the transverse momentum and invariant
mass of leptons. We have also excluded the some region of the
invariant mass distribution near $Z$-boson mass to suppress the
reducible background arises from $WZ$ production.      
Table \ref{signal_3l} also 
include signal (for different values of $R^{-1}$) and SM background
cross-sections after the selection cuts ($p_T$ cuts and cuts on the
invariant mass of the three lepton pairs). Here it is important to
mention that the transverse mass distributions of the three background
leptons also show a characteristics $W$-boson peak. However, we found
that the cuts on the invariant masses are more effective than the cuts
on the transverse masses. Numbers presented in Table
\ref{signal_3l} clearly indicates that, at $100~fb^{-1}$ luminosity of
the LHC, $5\sigma$ discovery of the {\em 3-lepton +
  $p_T\!\!\!\!\!\!/~~$} signal is possible upto $R^{-1}=600$ GeV.

\section{Signature of $(1,0)$-mode $U(1)$ sector at the ILC}

The importance of $\bmu$ production in association with a
$\bh$ has been briefly mentioned in the 
section 3. 
Since the couplings of $U(1)$ gauge boson or spinless adjoint
with quarks are suppressed by the hypercharge of the respective
quark, the $\bmu\bh$ pair production cross-section is
very small at LHC. Therefore, if not possible to study the signals
from $\bmu\bh$ pair production at the LHC.  
However, the couplings of $U(1)$ gauge boson or spinless
adjoint with leptons are enhanced by the hypercharges of the
corresponding leptons. One can expect a higher rate of production for
$U(1)$ gauge boson and spinless adjoint at a $e^+e^-$ collider. In
this section we will discuss the 
prospect of $\bmu\bh$ pair production at a future linear
$e^{+}e^{-}$ collider.   

\begin{figure}[t]

\begin{center}
\epsfig{file=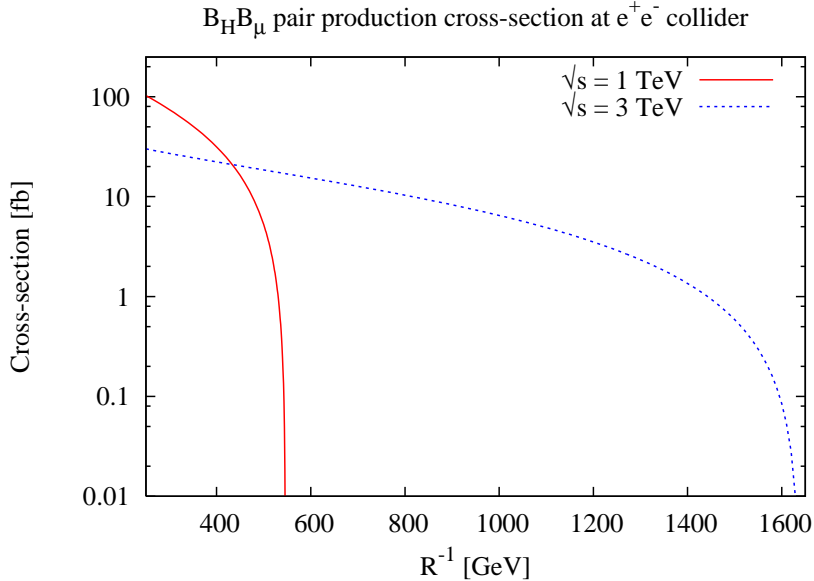,width=8cm,angle=270}
\end{center}
\caption{$\bmu\bh$ pair production cross section as a
  function of the compactification radius $R^{-1}$ for 1 TeV (solid
  line) and 3 TeV (dashed line) center-of-mass energy of $e^{+}e^{-}$
  collider.}
\label{cross_ILC}

\end{figure}

Both $\bmu$ and $\bh$ can couple to a $(1,0)$-mode
lepton and a SM lepton. The couplings arise from the compactification
of 6D kinetic term for 6D leptons and can be found in Appendix A. At an 
$e^{+}e^{-}$ collider, they can be directly produced in pairs,
\be
e^{+} + e^{-}~\to~\bmu + \bh,
\ee
which proceeds via the exchange of an $(1,0)$-mode electron
($E_{+}^{(1,0)}$ or 
$E_{-}^{(1,0)}$) in $t~(u)$ channel. 
The numerical values of the
cross-sections are presented in Fig. \ref{cross_ILC} as a function of
$R^{-1}$ for two different values of $e^{+}e^{-}$ center-of-mass
energies.


$\bmu$ dominantly decays to two SM leptons and a LKP resulting into
2-leptons and missing energy signal. However,
this is not the only source of 2-lepton and missing energy
in 2UED. Pair productions of almost all combinations of electroweak
gauge bosons and spinless adjoints give rise to 2-lepton and
missing energy final state. In fact, the largest contribution to 2-lepton and
missing energy, in 2UED, comes from the pair production of
$(1,0)$-mode electrons and their subsequent decays to SM electrons and
LKP \cite{kong}. However, in this work we will concentrate on the other
decay mode of $\bmu$, i.e to a photon and LKP. We are
interested in analysing the $\gamma+E\!\!\!/$ signal. 

{\em Single photon + missing energy} signal is
particularly interesting because this signal is the characteristics of
this theory. Since the structure of 2UED theory and 1UED theory
are almost identical, they give rise to
similar kind of signals at colliders. However, {\em single photon +
  missing energy} signal 
is one of the very few spectacular signals which can distinguish 1UED
and 2UED. In 1UED similar kind of signal can only arise from the
production of LKP in association with a photon. In fact, all those
theories which include a dark matter candidate 
can give rise to {\em single photon + missing transverse momentum}
signal\footnote{It is important to mention that in 
  MSSM the production of lightest 
nutralino ($\tilde \chi_{1}$) in association with the next to lightest
nutralino ($\tilde \chi_{2}$) and subsequent radiative decay of
$\tilde \chi_{2}$ to a photon and $\tilde \chi_{1}$ \cite{nutralino}
gives rise to the 
similar kind of signal. It is beyond the scope of this article to
compare this process with 2UED. 
}
via the radiative production of the dark matter candidate. As for
example, radiative production of lightest supersymmetric particles
(LSP) in different R-parity conserving supersymmetric models gives
rise to single photon plus missing transverse momentum signal. However, all
those radiative production processes are suppressed by the square of
an additional electron photon coupling. Moreover, photons from
radiative productions are predominantly soft or collinear. Therefore, the
detector acceptance cuts on photon energy and rapidity almost remove
the contributions from radiative pair production of LKP or LSP.

Radiative neutrino production ($e^{+}e^{-}~\to~\nu_{l}\bar
\nu_{l}\gamma$) is the major SM background to the signal.
Some acceptance cuts on the signal and background are listed in Table
\ref{cut_acc_ILC}. It is worthwhile to mention that photon rapidity
cut reduces another potentially 
dangerous background namely the radiative Bhabha scattering, $e^{+}e^{-}\to
e^{+}e^{-}\gamma$, where both the final state leptons escape along the
beam pipe.  

\begin{table}[h]

\begin{center}

\begin{tabular}{c|c|c}
\hline\hline

Kinematic Variable & Minimum value & Maximum value \\\hline\hline
$\eta_{\gamma}$ & $-$2.5  & 2.5 \\
$E_{\gamma}$     & 10 GeV   &  -  \\\hline \hline

\end{tabular}

\end{center}

\caption{Acceptance cuts on the photon rapidity
  $\eta_{\gamma}$ and photon energy $E_{\gamma}$}

\label{cut_acc_ILC}
\end{table}

We have estimated background cross-sections for two center-of-mass
energies of $e^{+}e^{-}$ collider. In
Table \ref{s_b_ILC}, we have presented the numerical values of the signal
and background  cross-sections for three different values of $R^{-1}$
at 1 TeV and 3 TeV $e^{+}e^{-}$ collider. Table \ref{s_b_ILC} shows
that the signal cross-sections (at 1 TeV) are small in comparison with the
background for the values of $R^{-1}$ close to the
kinematic limit of the collider. The situation is even worse for 3 TeV
collider. Here the signal is much smaller compared to the  background
even for the smaller value of $R^{-1}$.  


\begin{table}[h]

\begin{center}

\begin{tabular}{|c|c|c||c|c|c|}
\hline \hline
\multicolumn{3}{|c||}{$\sqrt{s}~=~1$ TeV} &
\multicolumn{3}{|c|}{$\sqrt{s}~=~3$ TeV}
\\\hline 
$ R^{-1}$ & Signal & Background & $R^{-1}$ & Signal & Background \\
in GeV  & in fb  & in fb & in GeV  & in fb  & in fb \\\hline\hline
300     & 26.03 & & 500 & 6.21 & \\ 
400     & 11.15 & 3609 & 1000 & 2.28 & 4248\\ 
500     & 1.87 & & 1500 & 0.21 & \\\hline\hline 
\end{tabular}

\end{center}

\caption{Signal and background cross-section at 1 TeV and 3 TeV
  collider after imposing acceptance cuts on the photon rapidity and
  energy.}

\label{s_b_ILC}
\end{table}

     
Pair production of $\bh$ in association with a photon also
gives rise to {\em single photon + missing energy} signal.
\be
e^{+} + e^{-}~\to~\gamma + \bh + \bh
\ee
However, cross-section of this process is suppressed w.r.t. the former. Also
the photons, 
radiated from incident electron or positron, are predominantly soft or
collinear. So the cuts on photon rapidity and energy completely
remove this contribution. We have estimated $\gamma \bh \bh$
contribution at $R^{-1}=500$ GeV and it is found that this is
suppressed by a large factor ($\sim 200$ at an 1 TeV collider) compared
to the $\bmu \bh$ contribution.

\begin{figure}[t]

\begin{center}
\epsfig{file=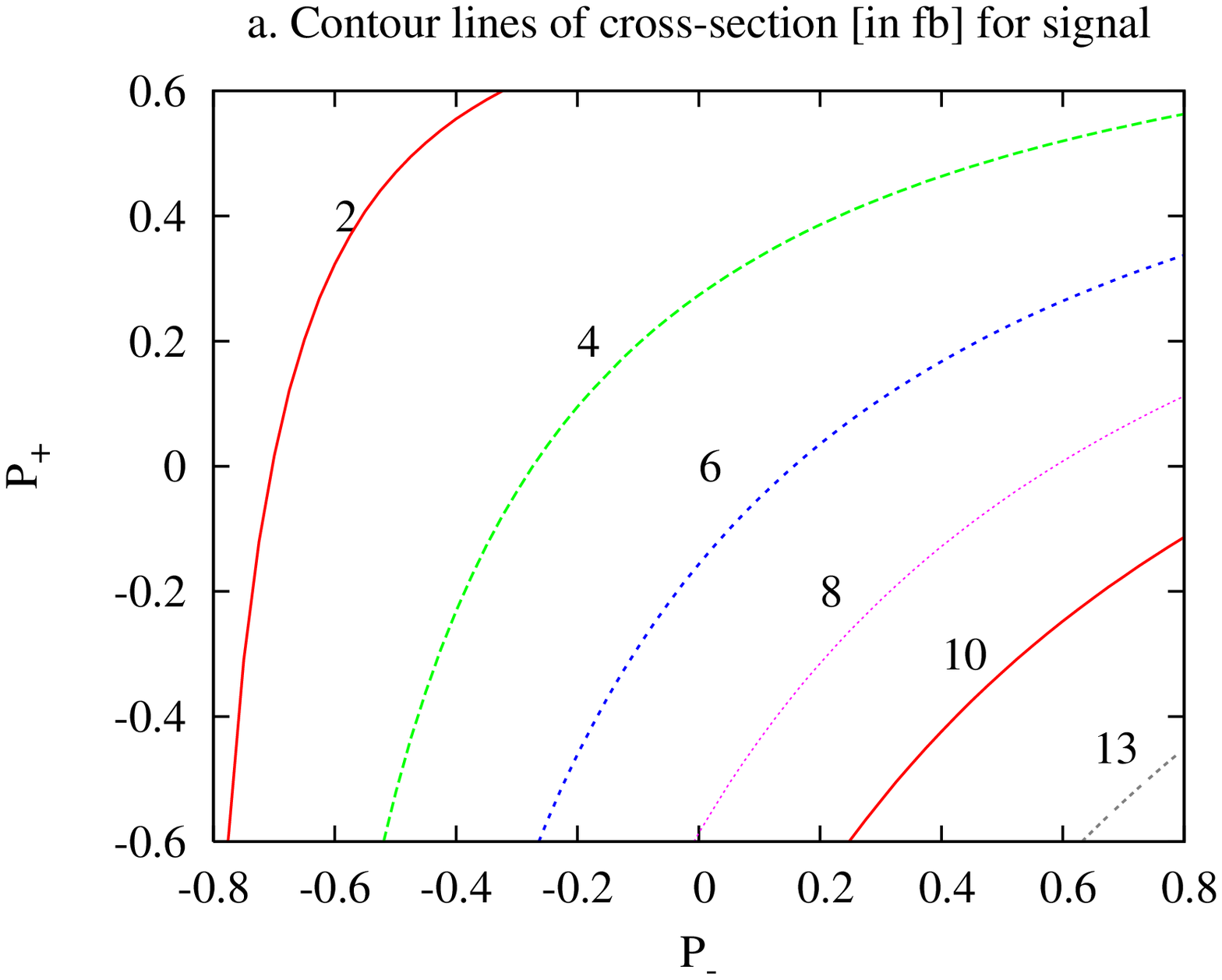,width=6.5cm,angle=270}
\epsfig{file=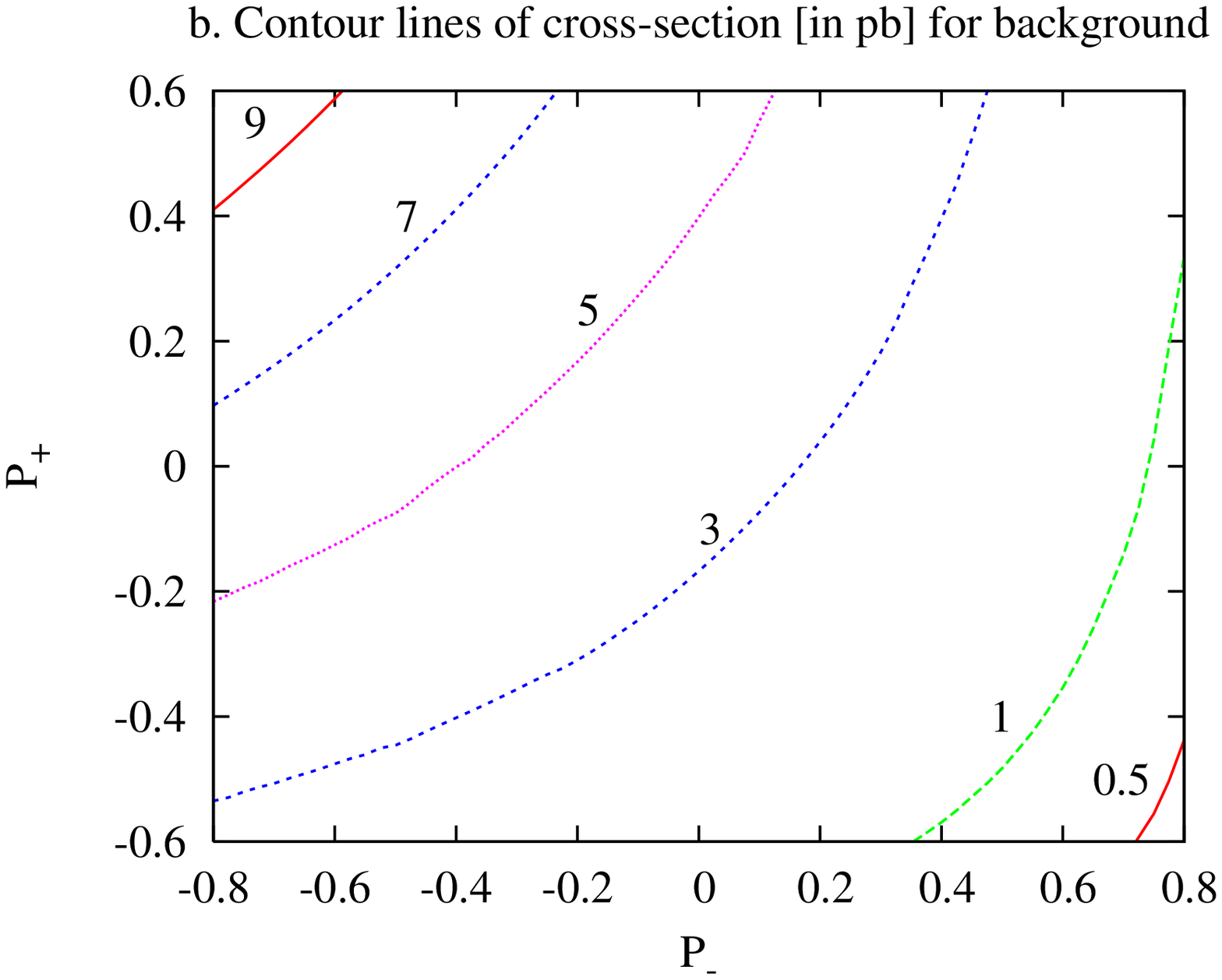,width=6.5cm,angle=270}
\end{center}
\caption{The beam polarization dependence of (a) $\sigma(e^+e^- \to
  \bmu\bh)$ for $R^{-1}=500$ GeV and (b) SM
  background at 1 TeV center-of-mass energy of $e^{+}e^{-}$ collider.}
\label{contur_ILC}

\end{figure}

\subsection{Beam polarization dependence of signal and background}

One of the merits for an $e^{+}e^{-}$ collider is the possibility of
highly polarized $e^+$ and $e^{-}$ beams. Here it is important to mention that
the maximum 80\% longitudinal beam polarization of electron beam and
60\% longitudinal beam polarization of positron beam is possible at
the ILC and CILC \cite{coll_phys}. Denoting the average $e^{\pm}$ beam
polarization by 
$P_{\pm}$, the polarized squared matrix element can be constructed
\cite{coll_phys} based on the helicity amplitude ${\cal
  M}_{\sigma_{e^+}\sigma_{e^-}}$:
\bea
\overline{\sum |{\cal M}|^{2}}&=&\frac{1}{4}[(1-P_{-})(1-P_{+})|{\cal
    M}_{LL}|^2+(1-P_{-})(1+P_{+})|{\cal M}_{LR}|^2\nonumber \\ 
&+&(1+P_{-})(1-P_{+})|{\cal M}_{RL}|^2+(1+P_{-})(1+P_{+})|{\cal
    M}_{RR}|^2]
\label{pol_m2}
\eea
Eq. (\ref{pol_m2}) clearly indicates that $P_{\pm}=1$ corresponds to
purely right handed and $P_{\pm}=-1$ corresponds to purely left handed
electron or positron beam. Since the electroweak interactions of both SM
and 2UED are chiral, it may be possible to suppress SM background
compared to the 2UED signal by proper choice of beam polarization.

Since electron is nearly massless, the background amplitude vanishes unless
electron and positron have opposite helicity (${\cal M}_{LL}^{B}={\cal
  M}_{RR}^{B}=0$). Production of $\bmu$ in association with a
$\bh$ is mediated by a massive $(1,0)$-mode electron
($E_{\pm}^{(1,0)}$) exchanged in $t~(u)$-channel. However, the structure of
2UED is such that $E_{+}^{(1,0)}~(E_{-}^{(1,0)})$ can only couple to a
left (right) handed electron or a right (left) handed positron and a
$(1,0)$-mode gauge boson or spinless adjoint. Therefore, here also the
amplitude
vanishes for the same helicity of electron and positron beam.
Since the hypercharge of right handed leptons are larger than left
handed leptons by a factor of 2, 
the coupling of a right (left) handed electron (positron) with
$E_{-}^{(1.0)}$ and $\bh^{(1,0)}$ or $\bmu^{(1,0)}$ is enhanced by a
factor two than the coupling of a left (right) handed electron
(positron) with $E_{+}^{(1.0)}$ and $\bh^{(1,0)}$ or
$\bmu^{(1,0)}$. Thus the contribution from $E_{-}^{(1,0)}$ exchange to
the cross-section is about a factor 16 larger than the $E_{+}^{(1,0)}$
contribution. The background process mainly proceeds via $W$ boson
exchange. Thus positive electron beam polarization, $P_-$, and negative
positron beam polarization, $P_+$, enhance signal cross-section and
reduce background at the same time. We have presented the beam
polarization dependence of the cross-section $\sigma(e^+e^- \to
\bmu^{(1,0)}\bh^{(1,0)})$ for $R^{-1}=500$ GeV and $\sqrt s=1$ TeV in
Fig. \ref{contur_ILC}a. Fig. \ref{contur_ILC}b shows the polarization
dependence of background cross-section $\sigma(e^+e^- \to \gamma
\nu_l\bar\nu_l)$ at 1 TeV center-of-mass energy of $e^+e^-$ collider.       

\begin{figure}[t]

\begin{center}
\epsfig{file=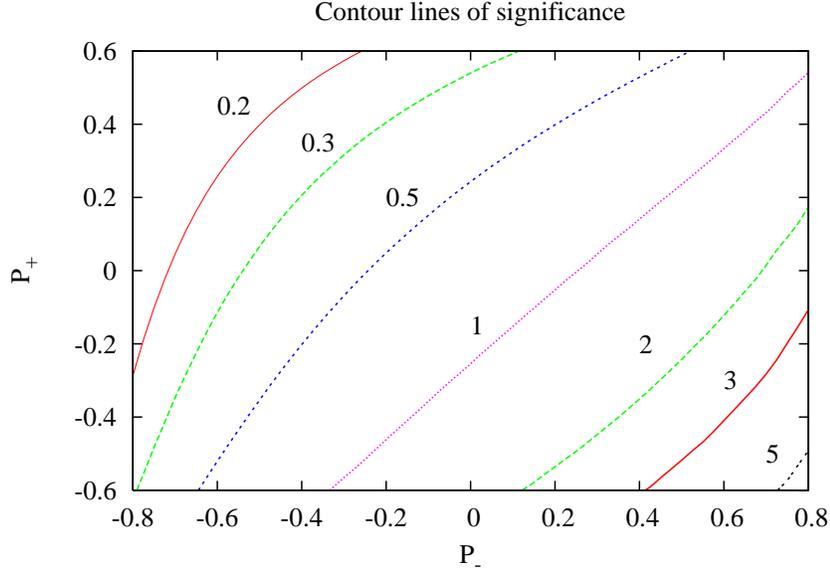,width=8cm,angle=270}
\end{center}
\caption{Contour lines of the significance for $e^+e^- \to
  \bmu\bh$ at $R^{-1}=500$ GeV, $\sqrt s=1$ TeV and
  ${\cal L}=500~fb^{-1}$.}
\label{signi_ILC}

\end{figure} 

We have used the same definition of the statistical significance as
defined in the Section 3.1. Fig. \ref{signi_ILC} shows the polarization
dependence of the significance for $\gamma+E_T\!\!\!\!\!\!/~~$ signal
at $R^{-1}=500$ GeV, ${\cal 
L}=500~fb^{-1}$ and $\sqrt {s_{ee}}=1$ TeV. In going from unpolarized beam
$(P_-,P_+)=(0,0)$ to partially polarized beam $(P_-,P_+)=(0.8,-0.6)$,
the significance is enhanced by a factor $\sim~8$ to $S=6$. But the
situation is not that hopeful for 3 TeV collider. At $R^{-1}=1.5$
TeV the significance enhanced by a factor $\sim 10$ to $S=0.7$ as we
vary beam polarization from $(P_-,P_+)=(0,0)$ to
$(P_-,P_+)=(0.8,-0.6)$.

\begin{figure}[t]

\begin{center}
\epsfig{file=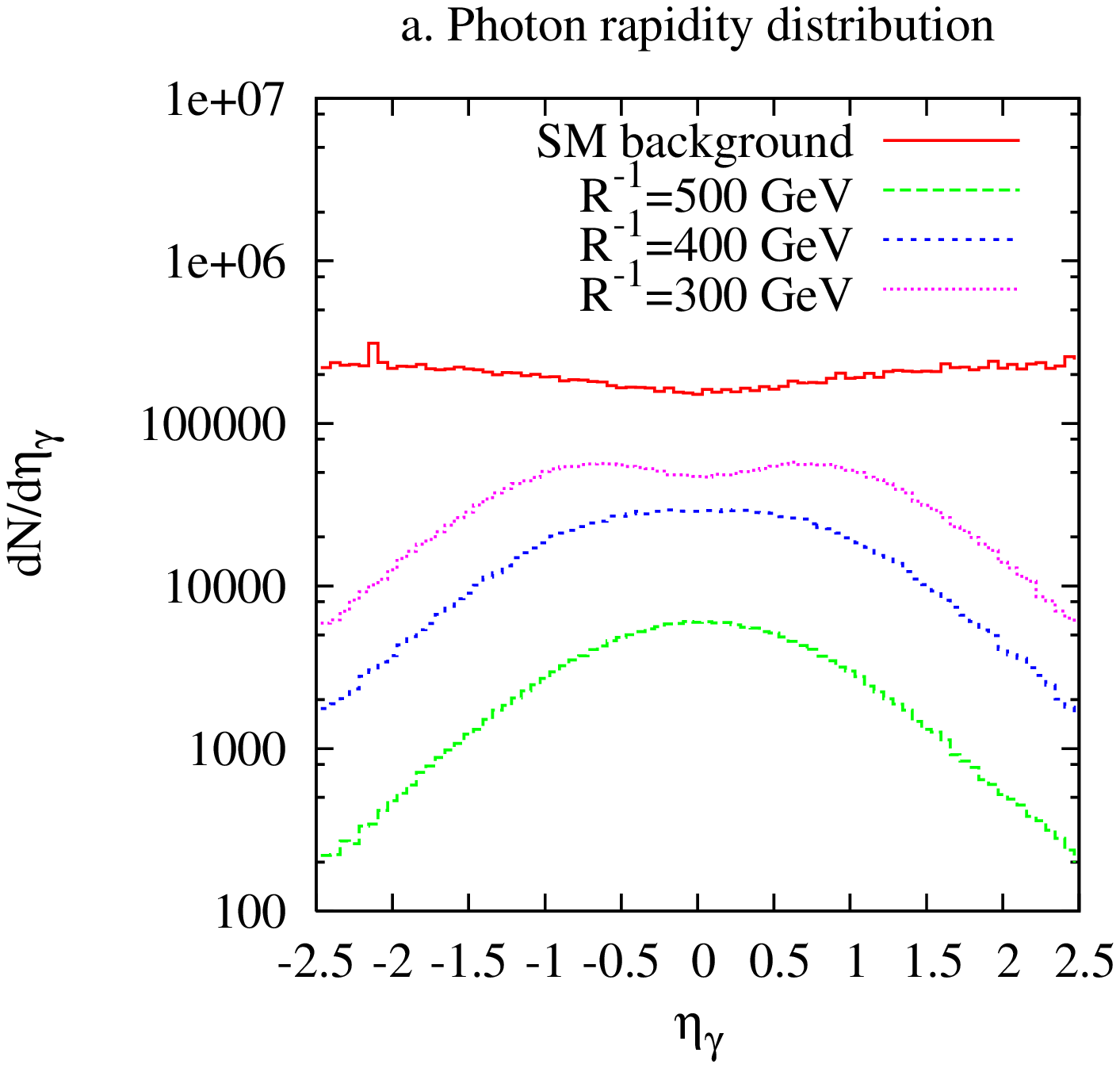,width=7.6cm,angle=270}
\epsfig{file=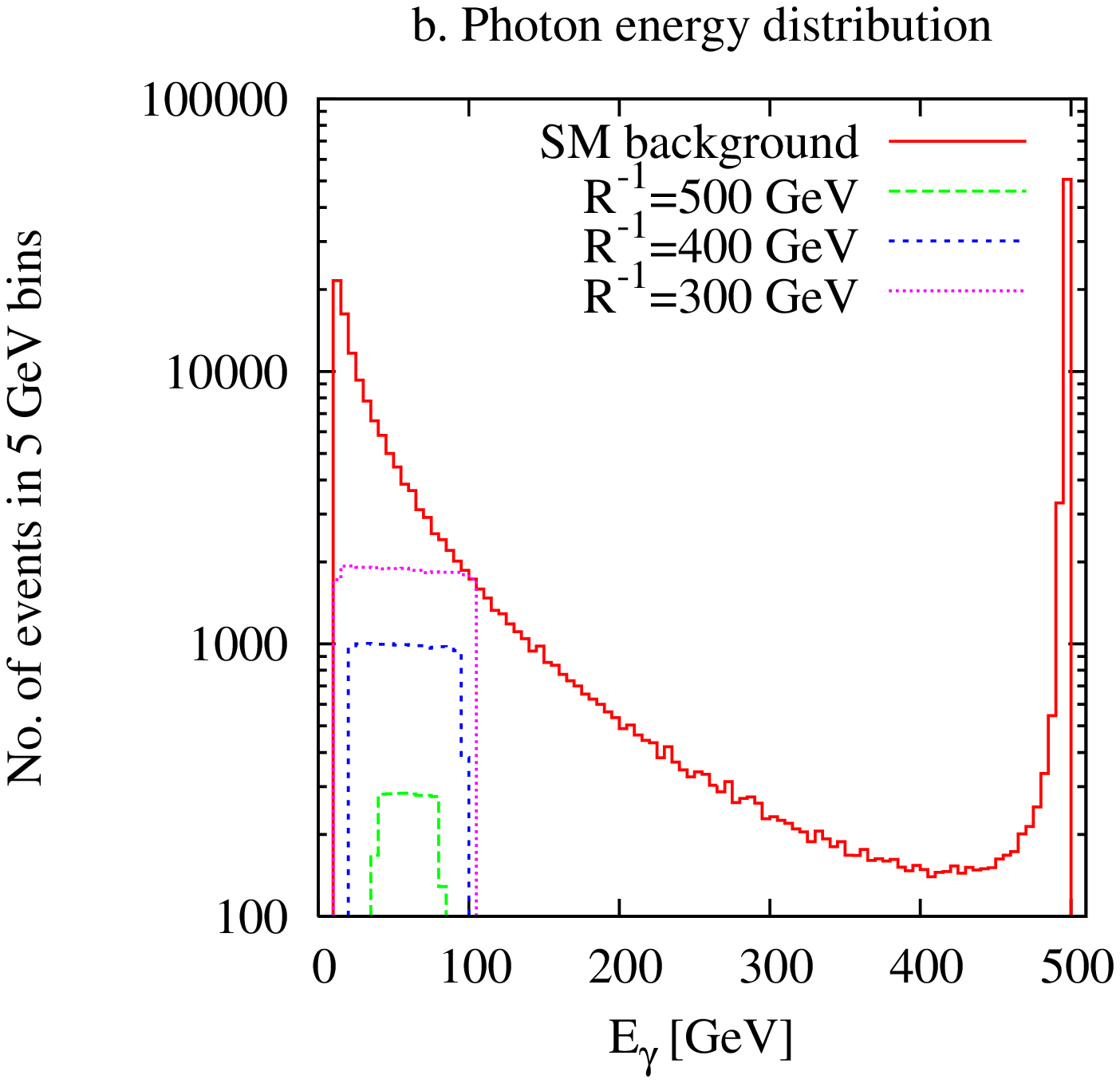,width=7.6cm,angle=270}
\end{center}
\caption{Photon (a) rapidity distribution and (b) energy distribution
  for $\gamma+E\!\!\!\!/~~$ events for signal (dotted, dashed,
  long dashed histogram) and background (solid histogram) at $\sqrt
  s=1$ TeV, ${\cal L}=500~fb^{-1}$ and beam polarization
  $(P_-,P_+)=(0.8,-0.6)$.} 
\label{dist_ILC}

\end{figure}

After exploiting the beam polarization to enhance the signal, further
suppression of background can be possible by looking at some kinematic
distributions. In Fig. \ref{dist_ILC}a we have presented the photon rapidity
($\eta_{\gamma}$) distribution for signal and background for $\sqrt{
s_{ee}}~=~1$ TeV. Fig. \ref{dist_ILC}a shows that signal
events are dominantly in the central rapidity region. Therefore, the cut on
photon rapidity (in Table \ref{cut_acc_ILC})
suppress the SM background compared to the signal. 

Since $W$-boson can only couple with left (right) handed electrons
(positrons), the selection of beam polarization, $P_-=0.8,~P_+=-0.6$,
reduces the background via t-channel $W$-boson exchange. Due to the
hypercharge, coupling of $Z$-boson with right (left) handed electrons
(positrons) is larger than left (right) handed electrons
(positrons). So the background from radiative production of
$Z$-boson increases as we select positive electron beam polarization
and negative positron beam polarization. However, radiative production of
$Z$-boson is characterized by a peak in photon energy distribution at
$E_{\gamma} = \frac{s-m_{Z}^{2}}{2\sqrt s}$. Fig. \ref{dist_ILC}b shows
$E_{\gamma}$ distribution of signal and background at an 1 TeV
center-of-mass energy of $e^+e^-$ collider for the previous choice of
$P_-$ and $P_+$ and
an integrated luminosity of ${\cal L}=500~fb^{-1}$. 

We will first concentrate on the $E_{\gamma}$ distribution of the
signal. The decay
$B_{\mu}\to \gamma \bh$ gives rise to a monoenergetic
photon with energy $E^{0}_{\gamma}=
\frac{m_{\bmu}^2-m_{\bh}^2}{2m_{\bmu}}$ at the
rest frame of $\bmu$. Photon energy spreads out due to the
velocity of  the produced $\bmu$, resulting into a box shaped
distribution. In a fixed center-of-mass energy $e^+e^-$ collider, as
we increase $R^{-1}$, $\bmu\bh$ pair production
approaches towards the kinematical threshold of the
collider. Therefore, increasing $R^{-1}$ implies decreasing velocity of
$\bmu$ and decreasing width of box shaped $E_{\gamma}$
distribution. 


\begin{table}[h]

\begin{center}

\begin{tabular}{|c|c|c||c|c|c|}
\hline \hline
\multicolumn{3}{|c||}{$\sqrt {s_{ee}}~=~1$
  TeV} & \multicolumn{3}{|c|}{$\sqrt {s_{ee}}~=~3$ TeV} \\ \hline \hline
$R^{-1}$& Signal & Background & $R^{-1}$& Signal & Background \\
in GeV  & Events & Events     &in GeV  & Events & Events     \\ \hline
\hline 
 300 & 35291 & 114760 (339) &  500  &  8279 &  141378 (376) \\  
 350 & 24192 & 93193 (305)  &  800  &  4528 &  69964 (264) \\
 400 & 15141 & 75194 (274)  &  1000 &  2835 &  53056 (230) \\
 450 & 7851  & 61669 (248)  &  1200 &  1154 &  21485 (146) \\
 500 & 2538  & 40369 (201)  &  1250 &  755  &  12735 (113) \\
 520 & 999   & 25538 (160)  &  1300 &  484  &  8182  (90) \\ \hline \hline

\end{tabular}

\end{center}

\caption{Number of $\gamma + E\!\!\!\!/~~$ signal and SM
  background events for 
two values of $e^+e^-$ center-of-mass energy assuming 500 $fb^{-1}$ 
integrated luminosity. $1\sigma$ fluctuations of the background events are
also shown in the brackets.}

\label{s_b_cut}
\end{table}


In Table \ref{s_b_cut}, the total number of signal events are
presented for different values of $R^{-1}$ and for two center-of-mass
energies of $e^+e^-$ collider. In Table \ref{s_b_cut}, we have used the
following event selection criteria. In a particular bin and also in
one of its adjacent bins, if the number of events are greater than
the SM background events plus $1\sigma$ fluctuation of SM background
events, we
take the background subtracted events in that particular 
bin as the signal events. The total number of signal events are the
sum of signal events in the above mentioned bins. The sum of the
background events in those bins are also presented in Table
\ref{s_b_cut} with their $1\sigma$ fluctuation. 

Now we will discuss one more interesting feature of this signal. For a
fixed center-of-mass energy $e^+e^-$ collider, the width ($\Delta
E_{\gamma}$) of photon energy distribution is directly related to the
$R^{-1}$ by the relation
\be
0.0405~R^{-4}-3.361~ s~ R^{-2} +s^2 ~=~87.3439~ s~ (\Delta E_{\gamma})^2
\label{width}
\ee 
From the measurement of the width of photon energy distribution
experimentally, one can calculate $R^{-1}$ by solving
Eq. (\ref{width}). But, experimentally the measurement of width will be
challenging task. Particularly, the measurement of the lower kinematical end
point ($E_{\gamma,min}$) of photon energy distribution will be
difficult due to the huge SM background in the lower $E_{\gamma}$
region and detector limitation of measuring very low energy
photon. However, the upper kinematical end point, $E_{\gamma,max}$, of photon  
energy distribution can be relatively easily measured. $E_{\gamma,max}$ is
related with $R^{-1}$ by the following relation
\bea
&~&0.107\left( 0.0405~ R^{-4}-3.361~ s~ R^{-2} +
s^2\right)^{\frac{1}{2}}+\nonumber\\   
&~&0.0535\left( 0.0405~ R^{-4}+ 0.4026~s~ R^{-2} +
s^2\right)^{\frac{1}{2}}=\sqrt s~ E_{\gamma,max}
\label{edge}
\eea
With the measured value of $E_{\gamma,max}$, Eq. (\ref{edge}) can
again be solved numerically to estimate the value of $R^{-1}$.

\section{Conclusion}

To summarise, we have investigated possible signatures of
$(1,0)$-mode electroweak particles in the framework of 2UED model in
the context of LHC and also at a future $e^{+}e^{-}$
collider. KK-parity allows only the pair production of these
particles. Once they
are produced in pairs, they give rise to {\em multi lepton + missing
transverse momentum} signal. The only exception is
$\bmu\bh$ production which gives rise {\em single
photon + missing transverse momentum} signal. 

At the LHC we study {\em two} and {\em three (charged
SM) lepton + missing transverse momentum} signal. We have estimated
contributions to the signal from 
different combinations of $(1,0)$-mode gauge bosons and spinless
adjoints pair production at the LHC. We have also estimated the SM
background 
contributions to {\em two} and {\em three charged lepton + missing
  transverse momentum} signal. We
find that with $100~fb^{-1}$ luminosity of the LHC, {\em two (three)
  lepton + missing transverse momentum} signal from the
$(1,0)$-mode electroweak sector of 2UED is greater than $5\sigma$
standard deviation of the SM background upto $R^{-1}=400~(600)$ GeV.    

At $e^{+}e^{-}$ collider we only concentrate on $\bmu \bh$ production.
$\bmu\bh$ production
cross-section is large due to the large hypercharge of electron and
positron. We have only investigated {\em single photon + missing
  energy} signal at a future $e^{+}e^{-}$ collider. The SM
model contribution 
($e^{+}e^{-}\to \gamma \nu \bar 
\nu$) to the {\em single photon + missing energy} is huge. Kinematical
cuts (cuts on photon energy) 
can remove only those part of background which  arises from the
radiative production of $Z$-boson. The dominant contribution
to $e^{+}e^{-}\to \gamma \nu \bar \nu$ arises from  t-channel
$W$-boson exchange. Fortunately, the choice of
positive electron beam polarization and negative positron beam
polarization reduces t-channel $W$-boson exchange contribution to the
background and at the same time enhance the signal cross-section. In our
analysis, we choose 80\% positive electron beam polarization and 60\%
negative positron beam polarization ($(P_{-},P_{+})=(0.8,-0.6)$). We
find that for $500~fb^{-1}$ luminosity of $e^{+}e^{-}$ collider,
{\em single photon + missing energy} signal form
$e^{+}e^{-}\to \bmu\bh$ production and
$\bmu\to \gamma \bh$ decay is greater than $5\sigma$
standard deviation of the SM background, almost upto the kinematic
limit of the collider. Since the photon arises from the $\bmu$
decay, the kinematical endpoints of
photon energy distribution only depend on the center-of-mass
energy, $\sqrt {s_{ee}}$, of $e^{+}e^{-}$ collider and $R^{-1}$. So from the
experimental measurement of one of the kinematical end points
(preferably upper kinematical end point) of photon energy distribution,
one can estimate $R^{-1}$. We find very high sensitivity of $R^{-1}$
with the upper kinematical end point of photon energy
distribution. Therefore, precise determination of photon energy
at electromagnetic calorimeter will enhance the accuracy $R^{-1}$
estimation.

\noindent
{\bf {Acknowledgments}} KG acknowledges the support from
Council of Scientific and Industrial Research, Govt. of India
(Sanction No. 09/028(0675)/2006 EMR-1). 

\vskip 10pt

\renewcommand{\theequation}{A.{\arabic{equation}}}
\setcounter{equation}{0}

{\bf Appendix A : Relevant Feynman Rules} \\

\begin{figure}[h]
\epsfig{file=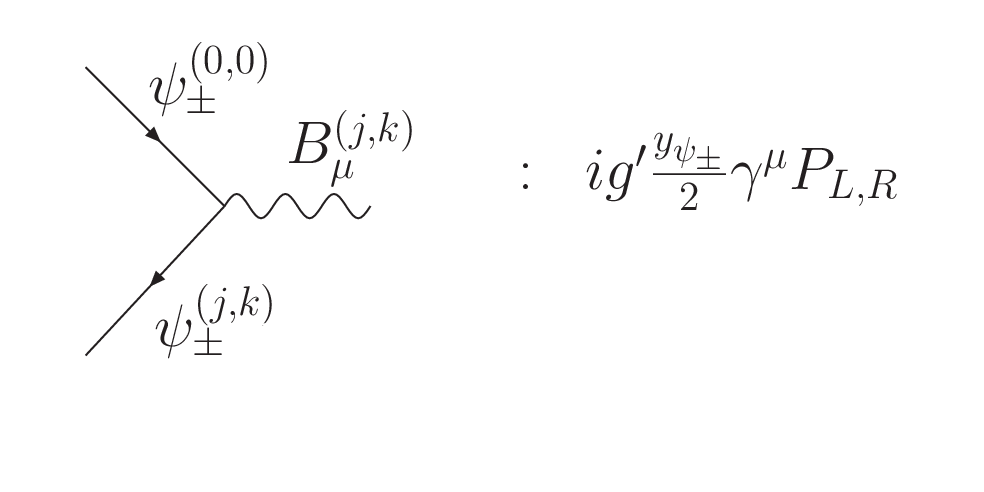,width=8cm}
\epsfig{file=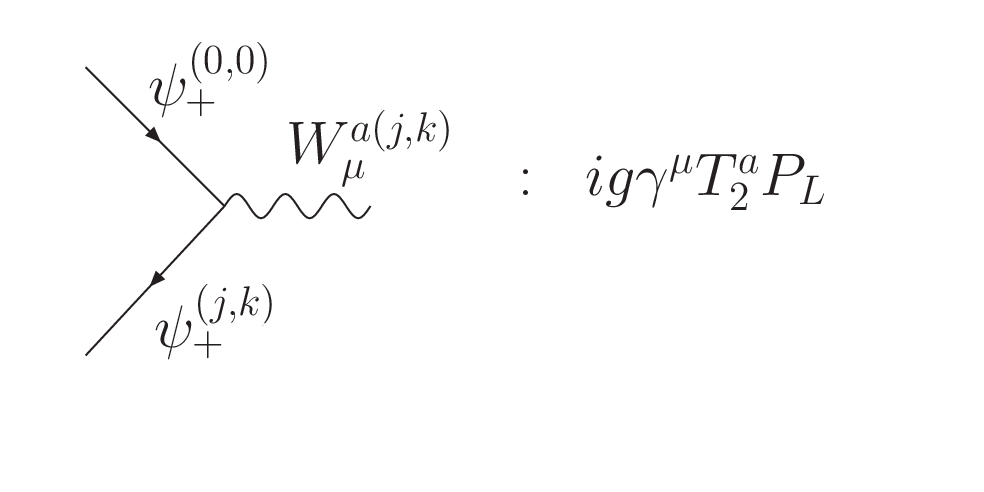,width=8cm}

\caption{Feynman rules of KK-number conserving interactions of a
  $(j,k)$-mode gauge boson ($B_\mu^{(j,k)}$ and $W_\mu^{3,\pm(j,k)}$) with
  the $(j,k)$-mode fermion and the corresponding SM fermion.  $g$ and
  $g^\prime$ are the $SU(2)$ and $U(1)$ gauge coupling constant
  respectively and $T^a_2$'s are the generators of the $SU(2)$ gauge
  group. $y_{\psi_\pm}$ is the hypercharge of the fermion $\psi_\pm$.}
\label{AQQ}
\end{figure}

\begin{figure}[h]
\epsfig{file=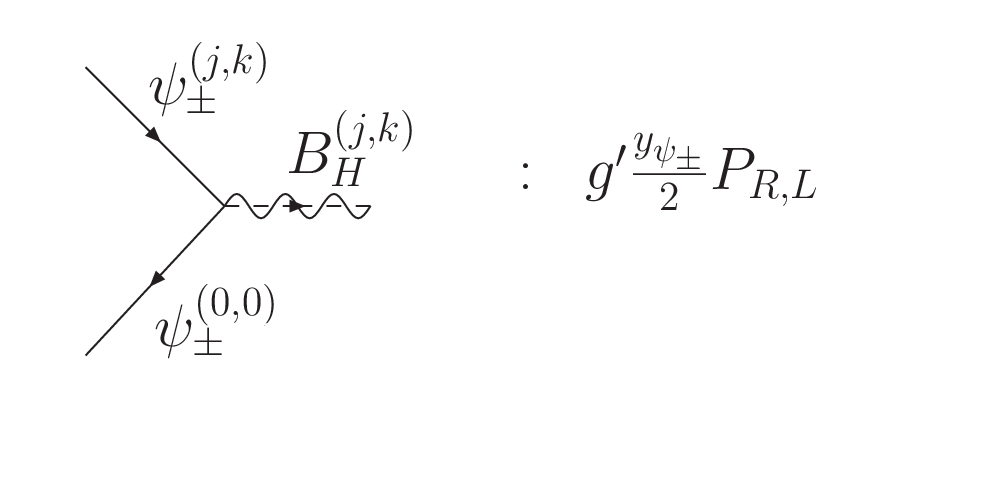,width=8cm}
\epsfig{file=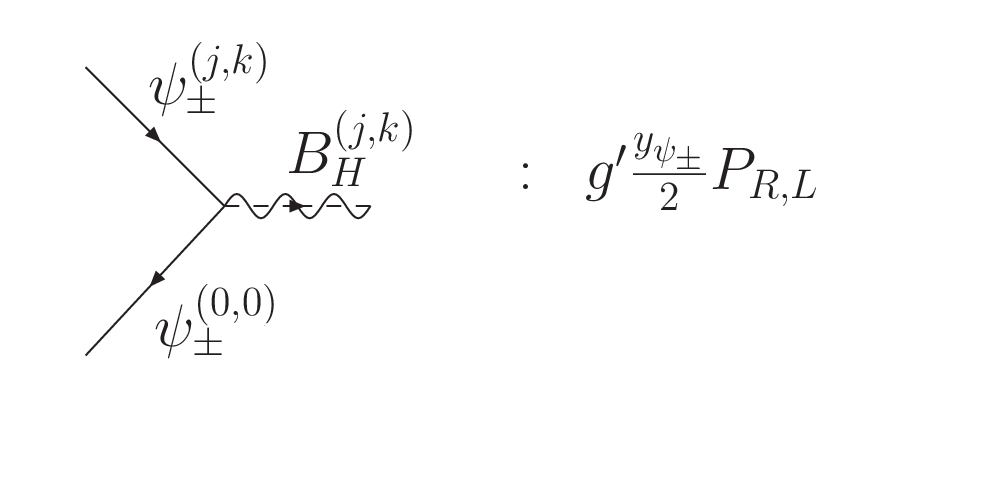,width=8cm}\\
\epsfig{file=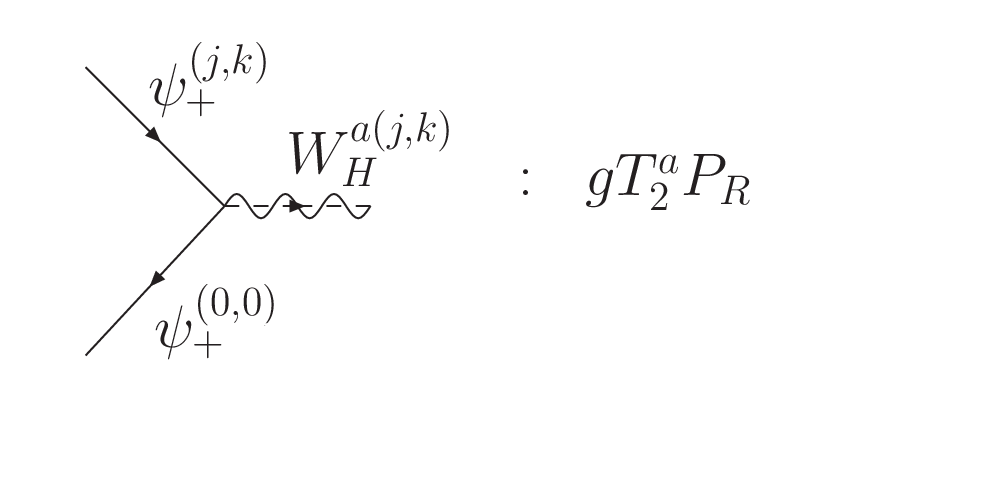,width=8cm}
\epsfig{file=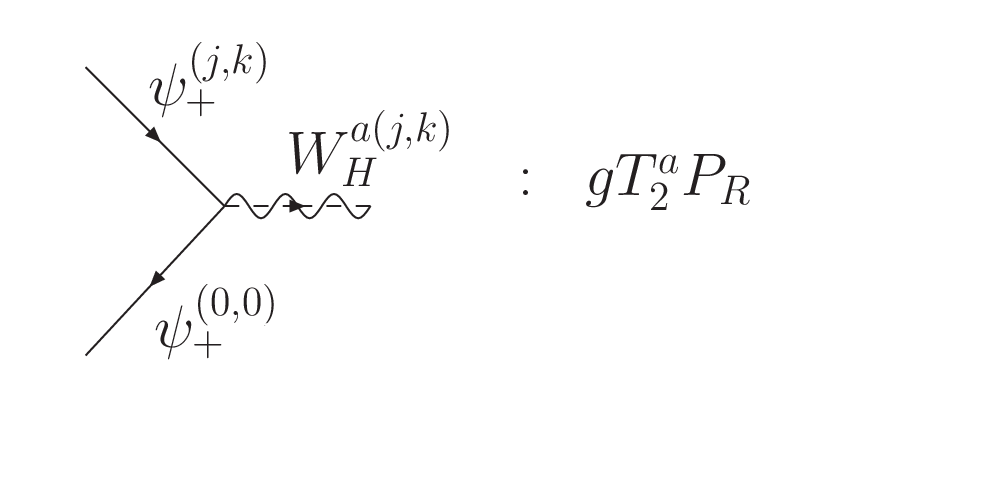,width=8cm}

\caption{Feynman rules of KK-number conserving interactions of a
  $(j,k)$-mode spinless adjoint ($B_H^{(j,k)}$ and $W_H^{3,\pm(j,k)}$) with
  the $(j,k)$-mode fermion and the corresponding SM fermion.}
\label{AhQQ}
\end{figure}

\begin{figure}[h]
\epsfig{file=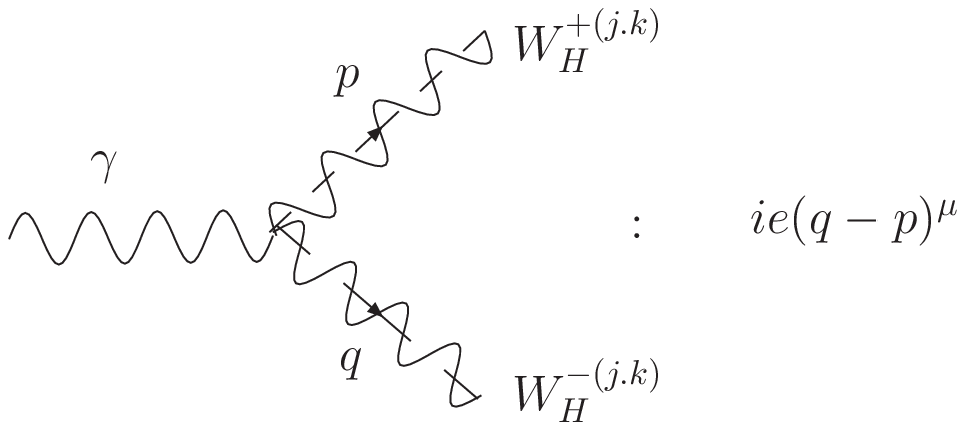,width=7cm}~~
\epsfig{file=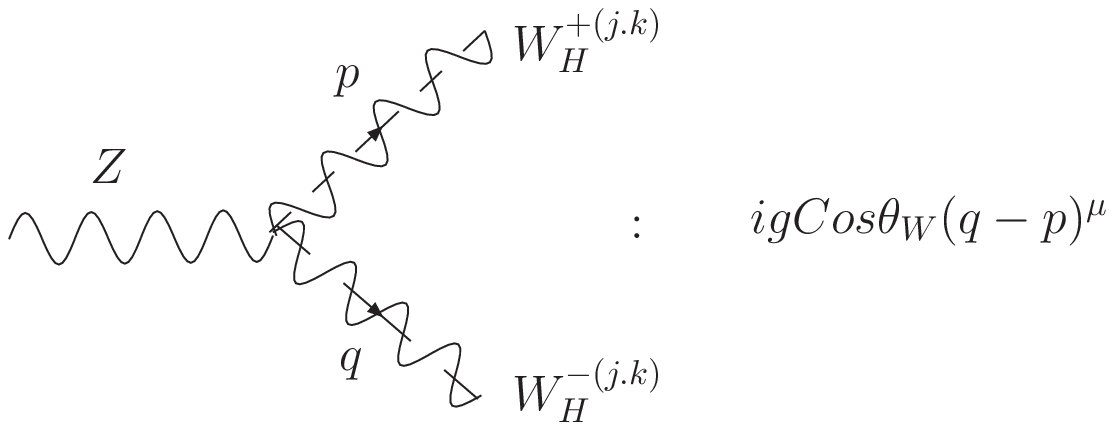,width=7cm}\\\\
\epsfig{file=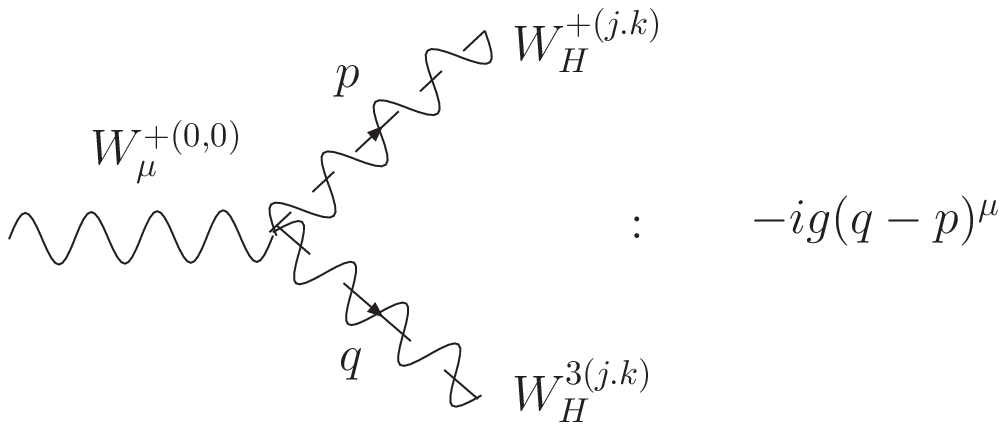,width=7cm}~~
\epsfig{file=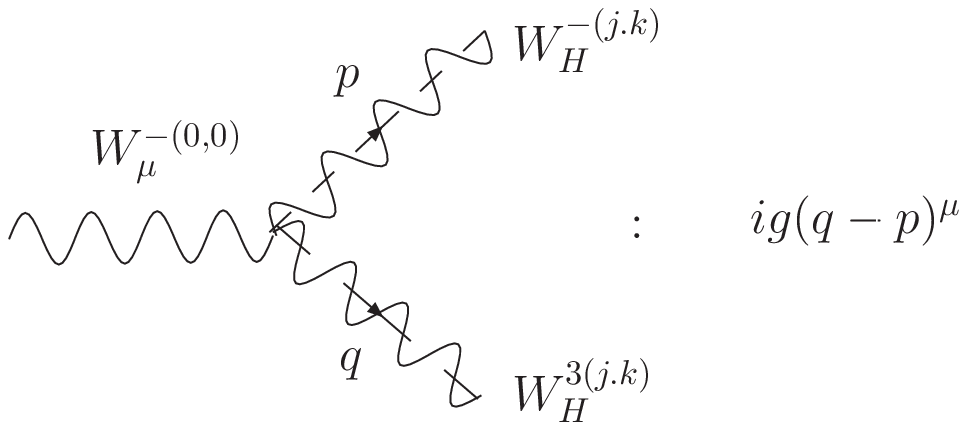,width=6.5cm}

\caption{Feynman rules of KK-number conserving interactions of two
  $(j,k)$-mode $SU(2)$ spinless adjoint ($W_{H}^{3,\pm(j,k)}$) with a
  SM vector boson ($\gamma,~Z$ or $W^{\pm(0,0)}_\mu$).}
\label{AWW}
\end{figure}

In this section we show the Feynman rules that are
relevant for the production of the $(1,0)$-mode electroweak particles
at the hadron collider and electron-positron collider. To make this
discussion more general, we present electroweak vertices
involving two $(j,k)$-mode particles and a SM particle. Corresponding
vertices involving $(1,0)$-mode can be easily inferred from the Feynman
rules given in Figs. \ref{AQQ}, \ref{AhQQ}, \ref{AWW}.

Compactification of the 6-dimensional kinetic terms for fermions and
integration over the compactified co-ordinates results KK-number
conserving interactions involving a KK gauge boson (spinless adjoint)
and two fermions. In Fig. \ref{AQQ}, we have presented $V^{(j,k)}_\mu
\psi^{(j,k)} \bar \psi^{(0,0)}$ ($V_\mu^{(j,k)}$ corresponds to
$B_{\mu}^{(j,k)}$ or $W_{\mu}^{3,\pm(j,k)}$) vertices and
Fig. \ref{AhQQ} shows $V^{(j,k)}_H
\psi^{(j,k)} \bar \psi^{(0,0)}$ ($V_H^{(j,k)}$ corresponds to
$B_{H}^{(j,k)}$ or $W_{H}^{3,\pm(j,k)}$) vertices. In Fig. \ref{AQQ} and
Fig. \ref{AhQQ}, $+$ and $-$ label the 6-dimensional chiralities of the
fermion as discussed in section 2 (see Eq. \ref{chirality}).

3-point interaction involving only one SM vector boson ($\gamma,~Z$ or
$W^{\pm}_\mu$) and two $(j,k)$-mode $SU(2)$ spinless adjoints
($W_{H}^{3,\pm(j,k)}$) arises from the
compactification of the self-interacting part of the 6-dimensional
$SU(2)$ gauge fields. Corresponding vertices are presented in
Fig. \ref{AWW}.

\begin{thebibliography}{99}

\bibitem{add} I.~Antoniadis,
  Phys.\ Lett.\ B {\bf 246}, 377 (1990); 
N.~Arkani-Hamed, S.~Dimopoulous and G.~Dvali, {
  Phys.~Lett.~}{B~}{\bf 429}, 263 (1998); I.~Antoniadis,
  N.~Arkani-Hamed, S.~Dimopoulos and G.~R.~Dvali, {
  Phys.~Lett.~}{B~}{\bf 436}, 257 (1998).

\bibitem{rs} L.~Randall and R.~Sundrum, { Phys.~Rev.~Lett.~}{\bf
83}, 3370 (1999); {\em ibid} {\bf 83}, 4690 (1999).

\bibitem{NPB550}
A.~Donini,~S.~Rigolin;
{ Nucl. Phys.} {\bf B550}, 59 (1999);
I. Antoniadis, K. Benakli, M. Quiros;
{Phys. Lett. B} {\bf 460}, 176 (1999).

\bibitem{PRD66}
C. Macesanu, C. D. McMullen, S. Nandi; 
Phys. Rev. D {\bf 66}, 015009 (2002);



\bibitem{PLB482}
A. De Rujula , A. Donini, M. B. Gavela, S. Rigolin;
Phys. Lett. B {\bf 482}, 195 (2000);
D. A. Dicus, C. D. McMullen, S. Nandi; 
Phys. Rev. D {\bf 65}, 076007 (2002);
C. Macesanu, C. D. McMullen, S. Nandi; 
Phys. Lett. B {\bf 546}, 253 (2002); 
C. Macesanu , A. Mitov , S. Nandi; 
Phys. Rev. D {\bf 68}, 084008 (2003);
C. Macesanu, S. Nandi, C. M. Rujoiu; 
Phys. Rev. D {\bf 73}, 076001 (2006).


\bibitem{UED} T.~Appelquist, H.~C.~Cheng and B.~A.~Dobrescu,
  { Phys. Rev. D}  {\bf 64}, 035002 (2001); 
H.~C.~Cheng, K.~T.~Matchev and M.~Schmaltz,
  { Phys. Rev. D} {\bf 66}, 056006 (2002).

\bibitem{unificUED}K. Dienes, E. Dudas, T. Gherghetta, { Nucl. Phys.} 
{\bf B537}, 47 (1999); K. Dienes, E. Dudas, T. Gherghetta, {
  Phys. Lett. B} 
{\bf 436}, 55 (1998); G. Bhattacharyya, A. Datta, S. K. Majee and
A. Raychaudhuri, { Nucl. Phys.} {\bf B760}, 117 (2007). 

\bibitem{dark_ued5}G. Servant, T. Tait, { Nucl. Phys.} 
{\bf B650}, 391 (2003); K. Kong. K. Matchev, {
  J. High Energy Phys.~}{\bf 038},~{0601},~(2006).

\bibitem{dark_ued6} B. Dobrescu, D.~Hooper, K.~Kong,
 R.~Mahbubani; {Jour. Cosmo. Astro. Phys.~}{\bf 0710}, 012 (2007).  

\bibitem{dobrescu} T. Appelquist, B. Dobrescu, E.~Ponton, H. Yee,
 { Phys. Rev. Lett.~}{\bf 87}, 181802 (2001).

\bibitem{dobrescu1} B. Dobrescu, E.~Poppitz,
 { Phys. Rev. Lett.~}{\bf 87}, 031801 (2001). 

\bibitem{dobrescu2}
 B. Dobrescu, E.~Ponton, { J. High Energy Phys.~}{\bf 071},~{0403},~(2004).

\bibitem{dobrescu4} B. Dobrescu, K.~Kong, R.~Mahbubani, { J. High
  Energy Phys.~} {\bf 006},~{0707},~(2007).

\bibitem{dobrescu3} G.~Burdman, B.~Dobrescu, E.~Ponton, {
Phys. Rev. D~}{\bf 74}, 075008 (2006).

\bibitem{KGAD1}
 K.~Ghosh and A.~Datta,
{ Nucl. Phys.} {\bf B800}, 109 (2008).

\bibitem{kong}
 A.~Freitas and K.~Kong,
  J. High
  Energy Phys. {\bf 0802}, 068 (2008).

\bibitem{KGAD2}
  K.~Ghosh and A.~Datta,
  Phys.\ Lett.\  B {\bf 665}, 369 (2008).

\bibitem{loop} E.~Ponton, L.~Wang, { J. High Energy Phys.~}{\bf 018},~{0611}, (2006).


\bibitem{PDG}
  C.~Amsler {\it et al.}  [Particle Data Group],
  Phys.\ Lett.\  B {\bf 667}, 1 (2008).



\bibitem{cteq4} H. Lai et al., {Phys. Rev. D~} {\bf 55}, 1280 (1997).

\bibitem{nutralino}
  S.~Ambrosanio and B.~Mele,
  Phys.\ Rev.\  D {\bf 53}, 2541 (1996).

\bibitem{coll_phys}
K. Hagiwara and D. Zeppenfeld, Nucl. Phys. {\bf B313}, 560 (1989).

\end{thebibliography}
\end{document}